\PassOptionsToPackage{numbers,sort&compress}{natbib}
\documentclass[onecolumn,preprintnumbers,number]{els-mrw}

\usepackage{amsmath,amssymb,amsfonts,amsthm,makeidx,graphicx}
\usepackage{txfonts}
\usepackage{helvet}
\usepackage{setspace}
\usepackage{xcolor}
\usepackage{natbib}

\makeatletter
\newcommand{\simpletoc}{%
  \begingroup
  \parindent=0pt
  \parskip=3pt
  \def\numberline##1{\makebox[3em][l]{##1}}%
  \def\contentsline##1##2##3##4{%
    \begingroup
    \def\toclevel{##1}%
    \def\chapterlevel{chapter}%
    \def\sectionlevel{section}%
    \def\subsectionlevel{subsection}%
    \ifx\toclevel\chapterlevel
      \textbf{##2}\dotfill ##3\par
    \else\ifx\toclevel\sectionlevel
      \hspace*{1em}##2\dotfill ##3\par
    \else\ifx\toclevel\subsectionlevel
      \hspace*{3em}##2\dotfill ##3\par
    \else
      \hspace*{5em}##2\dotfill ##3\par
    \fi\fi\fi
    \endgroup
  }%
  \@starttoc{toc}%
  \endgroup
}
\makeatother

\makeatletter
\renewcommand\normalsize{%
  \@setfontsize\normalsize{12pt}{14.4pt}%
}

\AtBeginDocument{%
  \normalsize
  \singlespacing
}

\begin{document}

\chapter{Quantum computing for high energy nuclear physics}\label{chap1}

\author[1]{Jo\~{a}o Barata}%

\address[1]{\orgname{CERN}, \orgdiv{Theoretical Physics Department}, \orgaddress{CH-1211, Geneva 23, Switzerland}}

\articletag{CERN-TH-2026-166}

\maketitle

\begin{abstract}[Abstract]
Quantum information science (QIS) offers a new set of conceptual and computational tools ideally suited to exploring the many-body, out-of-equilibrium, and real-time properties of strongly interacting quantum systems. These themes find a natural counterpart in high energy nuclear physics, whose central goal is to describe the properties of Quantum Chromodynamics (QCD), the theory of the strong interaction, under extreme conditions. Relevant examples range from the formation and internal structure of hadrons, to the collective and thermalization dynamics of strongly interacting QCD matter, and the properties of complex many-body scattering processes. Although tremendous progress has been made over the past decades in our understanding of QCD in the nuclear physics context, through perturbative methods, lattice QCD, and effective theory descriptions, phenomena that are intrinsically connected to real-time dynamics, highly entangled many-body evolution, or finite density regimes remain only partially accessible with existing approaches. In this article, we argue that QIS can help realize a new paradigm for addressing these open questions. With the aim of making the discussion accessible to both high energy nuclear physics and QIS audiences, we provide an overview of selected open problems in high energy nuclear physics, with emphasis on aspects motivated by collider phenomenology, and discuss how they can be approached using QIS tools. This connection is illustrated through representative examples drawn from the recent literature, making the case that QIS can become a driving force for discovery in nuclear physics in the coming years.
\end{abstract}

\simpletoc
\clearpage

\section{Introduction}\label{sec:intro}
High energy nuclear physics (NP) aims to understand how the strong interaction, described by Quantum Chromodynamics (QCD), defines the properties of ordinary matter across a wide range of physical regimes. In contrast to high energy particle physics, where the focus is often on understanding the elementary degrees of freedom and their short-distance interactions, NP is largely concerned with the collective, emergent, and many-body phenomena that arise in QCD. Paradigmatic examples include the internal structure of hadrons revealed in deep inelastic scattering~\cite{Gao:2017yyd, AbdulKhalek:2021gbh,Blumlein:2012bf}, the formation of quark-gluon plasma (QGP) in ultrarelativistic heavy-ion collisions~\cite{Busza:2018rrf,CMS:2024krd, Jacobs:2004qv}, and the behavior of strongly interacting matter at high temperature or large baryon density, as found in, e.g., the interior of neutron stars~\cite{Baym:2017whm,Vuorinen:2024qws,MUSES:2023hyz}. Over the past decades, developments along these different research lines have transformed our understanding of the strong force and of the Standard Model~\cite{Gross:2022hyw}.

\begin{figure}[h!]
    \centering
    \includegraphics[width=1\linewidth]{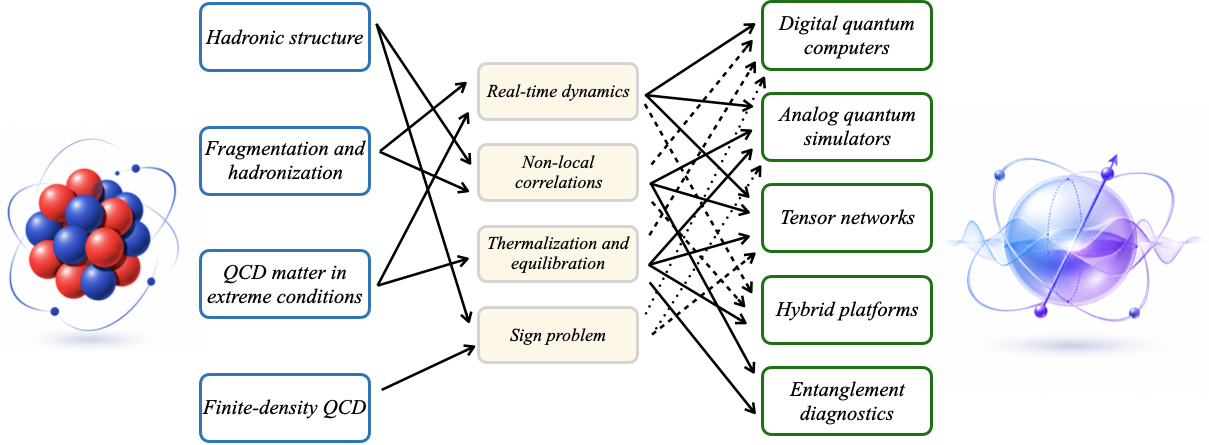}
    \caption{Schematic overview of the connection between selected open problems in high energy nuclear physics and the QIS tools. The arrows indicate representative connections: several NP questions involve multiple challenges, while different QIS methods can address more than one class of problems.}
    \label{fig:cartoon}
\end{figure}

Despite this remarkable progress, several central questions in high energy NP remain difficult to address using the standard theoretical toolkit. A prominent example is mapping the QCD phase diagram at finite temperature and baryon chemical potential. While Euclidean lattice QCD provides a first-principles non-perturbative formulation of the theory, the presence of a finite baryon chemical potential leads to a severe sign problem, preventing direct simulations in much of the high density regime~\cite{deForcrand:2009zkb,Fukushima:2010bq,Nagata:2021ugx}. Experimentally, this region has been explored through programs such as the Relativistic Heavy Ion Collider (RHIC) Beam Energy Scan, which searches for signatures of a phase boundary and a possible QCD critical point~\cite{STAR:2010vob,Kumar:2012fb}. However, even if experimental signatures of this structure were established, the absence of a direct first-principles theoretical counterpart would limit the extent to which they could be unambiguously interpreted. Similar limitations arise in a wide variety of problems involving real-time evolution, thermalization dynamics, confinement, and highly entangled many-body dynamics, where perturbative methods may fail, Euclidean lattice methods are not directly applicable, and effective descriptions are not always available. The path towards making significant progress in our understanding of these aspects of NP would therefore greatly benefit from the introduction of new tools that complement traditional methods. This motivation is summarized schematically in Fig.~\ref{fig:cartoon}, which organizes representative high energy NP problems (left column), matched to selected theoretical bottlenecks (central column).

In this article, we argue that promising tools to address these challenges can be found in the rapidly developing field of quantum information science (QIS), concerned with how quantum systems can be characterized, controlled, and used to process quantum information. This connection is indicated in Fig.~\ref{fig:cartoon} (right column), where we map the NP identified problems to the QIS toolkit. In the present context, its relevance stems from the possibility of engineering controllable quantum platforms, whose dynamics can be mapped onto quantum many-body models of interest. This idea was first envisioned by Feynman~\cite{Feynman:1981tf} and later formalized in the framework of universal quantum simulation~\cite{Lloyd:1996}. In recent decades, the practical realization of this program has undergone rapid development, driven by progress in digital quantum computing~\cite{Preskill:2018}, analog quantum simulation~\cite{Georgescu:2014}, and quantum-inspired classical methods~\cite{Schollwock:2011,Orus:2014,Banuls:2019bmf,Rudolph:2025gyq}. These developments have enabled controlled studies of, e.g., the real-time and out-of-equilibrium dynamics, of spin chains in regimes that are challenging for standard classical approaches~\cite{Martinez:2016,Banuls:2019bmf,Bauer:2022hpo,DiMeglio:2023fys}. On top of that, QIS has also provided new observables and diagnostics to probe the properties of quantum systems, enabling a more complete tomographic description of these~\cite{Elben:2022jvo,Amico:2007ag}. The central task at hand is therefore to translate the progress achieved in the simulation and characterization of controllable quantum mechanical systems to high energy NP. This article thus summarizes the current progress in realizing this connection.

Given the interdisciplinary nature of this article, the focus is put on introducing a selection of key concepts and open questions in high energy nuclear physics for which QIS methods may offer new insights, rather than providing exhaustive technical introductions to either field separately. For broader background introductions, we refer interested readers to recent reviews on high energy nuclear physics topics~\cite{Busza:2018rrf,Boussarie:2023izj, Accardi:2012qut,Morreale:2021pnn,Gelis:2010nm}, quantum computing and quantum simulation~\cite{Georgescu:2014,Jha:2023bcb,Gill:2024nfh,deWolf:2019lhx}, and the growing interface between high energy physics and QIS~\cite{DiMeglio:2023fys,Banuls:2019bmf,Halimeh:2025vvp}. Compared with previous broader scope reviews~\cite{Bauer:2022hpo,DiMeglio:2023fys,Halimeh:2025vvp}, the emphasis here is on the connection between QIS methods and questions motivated by high energy (nuclear) collider phenomenology, rather than focused on lattice gauge theories or more general aspects in high energy physics. The manuscript is organized as follows. Section~\ref{sec:Nuclear} introduces a selection of active research questions in high energy NP, emphasizing topics with theoretical, phenomenological, and experimental relevance. Section~\ref{sec:QIS} briefly introduces several methods and tools from QIS that enable the controlled study of quantum many-body systems. Section~\ref{sec:applciation_NP} presents representative applications of these methodologies to problems closely connected to the core interests of high energy nuclear physics. Finally, Section~\ref{sec:conclusion} summarizes the main points of the discussion and outlines possible future directions.

\section{Open problems in high energy nuclear physics}\label{sec:Nuclear}

In the following discussion we present some open problems in high energy nuclear physics. The discussion is mainly focused on topics relevant for collider physics, starting from a discussion on high energy scattering. The goal is not to present an extensive and complete discussion on NP as a whole, which extends beyond the context of collider physics, but rather provide motivating and explicit cases where QIS tools can be applied to provide novel physical insights.

\subsection{High energy scattering}\label{subsec:pp}
A lot of the progress and discoveries in high energy physics have been brought by the development of high energy collider experiments. Two paradigmatic examples of these are the RHIC at Brookhaven National Laboratory, which made the first experimental observation of a nearly ideal hydrodynamical quark gluon plasma state, and the Large Hadron Collider (LHC) at CERN, where the Higgs boson, the missing element of the Standard Model, was discovered. 

\begin{figure}[h!]
    \centering
    \includegraphics[width=1\textwidth]{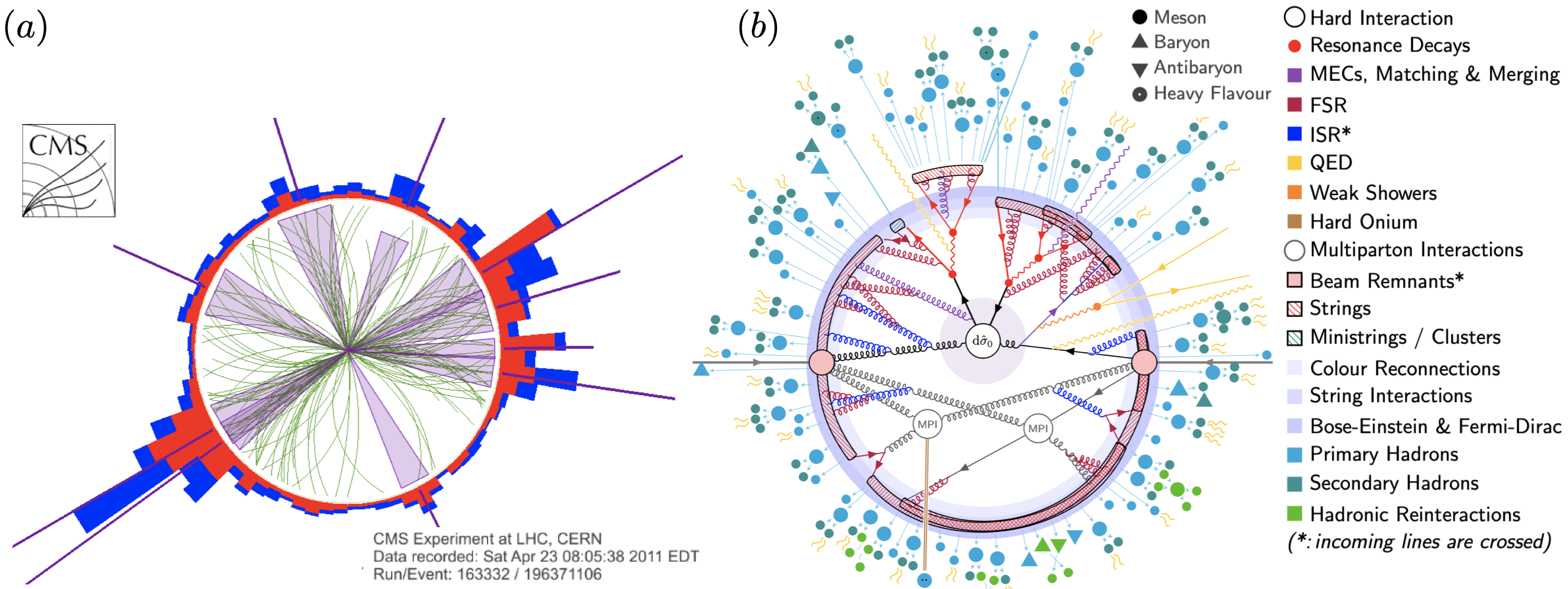}
    \caption{High energy scattering from experimental and theoretical perspectives. $(a)$ CMS proton-proton event display showing charged-particle tracks and calorimeter energy depositions. $(b)$ Schematic Monte Carlo description of a hadronic collision, from the hard interaction and parton shower to hadronization and decays. Figures adapted from~\cite{CMS:EXO-11-071,Bierlich:2022pfr}, respectively.}
    \label{fig:pp_fig}
\end{figure}

In general, in such experiments, particle packets, ranging from electrons to large hadrons, are boosted close to the speed of light inside circular accelerators, and made to scatter at particular interaction points. The outgoing state of such processes is measured by large detector set-ups placed around the interaction vertex. As these scattering processes involve large momentum exchanges, the measured final states are typically multi-particle states, as illustrated in Fig.~\ref{fig:pp_fig} $(a)$. There we can observe the tracks (lines inside the circle) and energy depositions (red and blue towers) of the produced particle states originating from the central interaction point in a proton-proton ($pp$) collision (protons move orthogonally to the page) at the LHC. Although different scattering processes lead to different final states, the common key goal is to infer the dynamical QCD processes that connect the simple initial scattering states to the complex multi-particle final distributions, from the underlying quantum field theory (QFT). We follow to describe how such a process can be understood from the theory point of view, why it is an intrinsically hard problem to solve, and its connection to the experimentally measured particle distributions shown in Fig.~\ref{fig:pp_fig} $(a)$.

High energy scattering processes are typically described under the umbrella of the so called $S$-matrix scattering picture, see e.g.~\cite{Lehmann:1954rq,Weinberg:1995mt}. In this framework, one assumes that the microscopic interactions take place only during a (short) finite time interval, while the initial and final states are well described by freely propagating particles far away from the collision region. The basic object of interest is therefore the transition amplitude between incoming (in) asymptotic states, and asymptotic outgoing (out) states. These are typically defined as
\begin{equation}
  |\alpha,{\rm in}\rangle
  =
  \lim_{t\to -\infty}
  U(0,t)\,e^{-iH_0t}\,
  |\alpha\rangle_{\rm free},
  \qquad
  |\beta,{\rm out}\rangle
  =
  \lim_{t\to +\infty}
  U(0,t)\,e^{-iH_0t}\,
  |\beta\rangle_{\rm free},
\end{equation}
where \(U(t_2,t_1)\) is the time-evolution operator in the interacting theory, i.e. we write the theory Hamiltonian as $H=H_0 +V$, with $H_0$ the free theory Hamiltonian. The states \(|\alpha\rangle_{\rm free}\), \(|\beta\rangle_{\rm free}\) are multiparticle
eigenstates of the free Hamiltonian $H_0$, but not of the full theory which includes interactions. The $S$-matrix is defined by the overlap between these states
\begin{equation}\label{eq:Smatrix}
  S_{\beta\alpha}
  =
  \langle \beta,{\rm out}|\alpha,{\rm in}\rangle
  =
  \lim_{T\to\infty} \, 
  _{\rm free}\langle \beta|\,e^{iH_0 T}\,
  U(T,-T)\,
  e^{iH_0 T}\,|\alpha\rangle_{\rm free} \, .
\end{equation}
In the second equality we have made explicit the form which can be more easily connected to the experimental program: the incoming states evolve under the interacting Hamiltonian over the interval $[-T,T]$, of duration $2T$, while the external free-evolution factors relate this evolution to the asymptotic incoming and outgoing states. Referring back to Fig.~\ref{fig:pp_fig} $(a)$, the initial nearly free states would be the incoming protons, while the experimentally measured states correspond to the asymptotically free outgoing particles.

Exactly determining Eq.~\eqref{eq:Smatrix} in a theory such as QCD is extremely non-trivial for several reasons. Firstly, QCD is a confining theory, meaning that the observable states are hadrons, i.e. color singlet states. However, at high energies these are not the \textit{natural} degrees of freedom to describe the theory, as this is best understood in terms of quarks and gluons (partons) degrees of freedom~\cite{Gross:1973id,Politzer:1973fx}, which are weakly interacting due to asymptotic freedom. Thus, while the incoming and outgoing states are well described by hadronic degrees of freedom, the hard interaction enclosed in $U(T,-T)$ is better formulated in terms of partons. As a result, the confinement mechanism, responsible for the hadronization of quarks and gluons into bound states makes the actual computation of the QCD $S$-matrix extremely non-trivial, requiring a description of this non-perturbative transition. A second important aspect -- even if one were able to accommodate both the partonic and hadronic pictures in a single framework (as in the Euclidean lattice formulation of QCD) -- the fully interacting operator in
\begin{align}
   U(T,-T) = e^{-i 2H \, T}\,  ,
\end{align}
requires to directly evolve the states in real-time. This points to a classic \textit{sign problem}, as the respective action would have a complex valued weight. This naively prevents it to be interpreted as a positive probability distribution, and thus the standard importance sampling techniques used in lattice QCD studies, can not be applied. Of course, as is commonly done in lattice studies~\cite{Rothe:1992nt}, one could Wick rotate time, making the action real valued. However, for real-time problems, the analytical continuation back to the Minkowski metric is ill-posed in a discretized lattice, and Euclidean simulations thus can not directly implement asymptotic real-time scattering, although scattering information can be extracted through finite-volume methods in suitable channels~\cite{Luscher:1990ck,Luscher:1990ux}.\footnote{ Extending this access to general multi-particle scattering remains challenging~\cite{Maiani:1990ca}. We note nonetheless, that there are other strategies which can give non-perturbative access to the QCD $S$-matrix, see e.g.~\cite{Roy:1971tc,Guerrieri:2018uew,Guerrieri:2020bto,Ananthanarayan:2000ht,Colangelo:2001df,He:2023lyy}, going beyond the scope of the present article.}

Instead, especially in the context of collider physics, scattering processes are mainly discussed within the realm of perturbative QCD. This approach is valid due to the above mentioned asymptotic freedom property, which ensures that at large energy transfers, QCD behaves as a nearly free partonic theory, and the corrections to this can be computed order by order in the strong coupling. This allows to perturbatively expand $U(T,-T)$, and provide systematically improvable predictions for the $S$-matrix. More, as the confinement/hadronization dynamics become more relevant at low energies, this suggests that the entire scattering process can be decomposed/\textit{factorized} into separate pieces characterized by well separated energy scales. The idea of factorization across energy scales, combined with the notion of \textit{universality}, i.e. that the underlying description in a given energy window is not exclusively tied to a particular underlying scattering process, are at the center of our understanding of high energy scattering in QCD. We follow to briefly discuss these notions in slightly more detail, emphasizing where current roadblocks exist, and refer a reader to~\cite{Collins:2011zzd,Ellis:1996mzs} for in-depth discussions on these topics.

Collider observables are not typically described by the complex valued transition amplitudes, but rather by the probabilities of observing certain transition processes between initial and final states, i.e. in terms of scattering cross-sections $\hat \sigma$. Scattering cross-sections are obtained by squaring the interacting elements of the $S$-matrix and averaging over the relevant quantum numbers~\cite{Schwartz:2014sze,Peskin:1995ev}:
\begin{equation}
  d\hat{\sigma}_{ab\to n}
  =
  \frac{1}{4\sqrt{(p_1\cdot p_2)^2-m_a^2m_b^2}}\,
  \left|
    \mathcal{M}_{ab\to n}
  \right|^2
  d\Phi_n .
\end{equation}
where $S_{\beta\alpha} = \delta_{\beta\alpha}+i(2\pi)^4\delta^{(4)}(p_\beta-p_\alpha)\mathcal{M}_{\beta\alpha}$, and $d\Phi_n$ is the relevant phase space factor for a two (incoming particle $a$ and $b$) to $n$ scattering process for massive particles. The matrix element $|\mathcal{M}_{\beta\alpha}|^2$ thus describes the probability associated to the different channels which are allowed in a two-particle scattering process. In what follows we assume that the scattering process occurs at large energies, characterized by a scale $Q$, set by the largest momentum exchange during the interaction. More, the hadronization transition becomes relevant only at the low-energy scale $\Lambda_{\rm QCD}$, roughly set by the inverse size of the proton. Thus, we work in the regime $Q \gg \Lambda_{\rm QCD}$, which effectively allows to separate the partonic dynamics from the confining sector. Indeed, employing this separation of scales, one can show the probability to measure a given observable $\hat O$ on the final state hadrons takes the (schematic) general form:
\begin{equation}\label{eq:factorization}
  d\sigma_{h_1h_2\to O}
  =
  \sum_{a,b}
  \int_0^1 dx_1
  \int_0^1 dx_2\,
  f_{a/h_1}(x_1,\mu_F)\,
  f_{b/h_2}(x_2,\mu_F)\,
  d\hat{\sigma}_{ab\to O}
  (x_1P_1,x_2P_2;\mu_F)
  +
  \mathcal{O}\left(
    \frac{\Lambda_{\rm QCD}^2}{Q^2}
  \right)\, ,
\end{equation}
if the process is sufficiently inclusive, see e.g.~\cite{Collins:2011zzd}. This relation significantly simplifies the calculation of the full cross-section. It separates the characterization of the initial state hadrons into a so called parton distribution function (PDF) \(f_{a/h}(x,\mu_F)\), which naively expresses the probability of finding a parton $a$ inside hadron $h$ carrying a longitudinal momentum fraction $x$ of the hadron momentum $P$. The cross-section $d\hat{\sigma}_{ab\to O}
  (x_1P_1,x_2P_2;\mu_F)$ thus contains the information of the now partonic scattering to some collection of final state hadrons, where the observable $\hat O$ is measured.\footnote{A similar idea can be applied to further separate the partonic scattering dynamics from the formation of the final state hadrons. This leads to a reciprocal form as found for the PDFs, and we refer an interested reader to~\cite{Collins:2011zzd} for further discussion.} Thus, the PDFs contain the low energy information about the structure  of the incoming hadrons, while perturbative methods can be used to explain the partonic scattering between their constituents. These processes are to be related at a factorization scale $\mu_F$, ensuring the consistency of the entire formulation.

This factorized picture underlies the modern understanding of high energy scattering processes in collider experiments, as illustrated in Fig.~\ref{fig:pp_fig} $(b)$, which should be contrasted with the experimental picture in panel $(a)$. Generalizing from the example discussed in Eq.~\eqref{eq:factorization}, one can separate the entire complex scattering process into a sequence of different physical mechanisms (indicated in the figure), which can be studied separately. The ones dominated by some large energy scale can be treated within a perturbatively driven approach, while others can only be computed using non-perturbative methods (such as the hadronization transition). Although at face value, this factorized picture does not seem to fundamentally solve the problem of reconstructing the entire $S$-matrix, as it still involves genuinely non-perturbative objects whose direct determination can remain challenging, it still maintains a certain predictive power. This is ensured by the idea of universality, by which the different objects entering the factorization formula, e.g. the PDFs, are common to a variety of scattering processes. This allows to constrain them in a data driven way, combined with the understanding of their scale evolution in perturbative QCD~\cite{Collins:2011zzd,Ellis:1996mzs}.

Despite the tremendous progress in the understanding of QCD encapsulated in Eq.~\eqref{eq:factorization} for collider phenomenology, it is clear that several aspects go beyond the reach of \textit{traditional} tools used in QCD, for example:
\begin{itemize}
    \item Can the validity of factorization formulas be fully tested?
    \item In a complementary way to the data driven extraction of PDFs and other non-perturbative objects, based on universality arguments, can one directly compute these objects from QCD theory?
    \item Can the cross-section $d \hat{\sigma}_{ab\to O}$ be understood in the limit of a large number of final state particles, where, e.g, collective dynamics might emerge, beyond the reach of the factorized form in Eq.~\eqref{eq:factorization}?
\end{itemize}
The following sections are dedicated to exploring some of these questions.

\subsection{Hadronic structure at high energies}\label{subsec:EIC}
In a high energy scattering event, the final state particles distributions are very much dependent on the properties of the incoming hadronic states. As a result, to make a connection between the properties of QCD and the observed final state particles, requires a precise and complete understanding of the non-perturbative internal structure of hadrons. Following the previous section, here we shall focus our discussion on perhaps the simplest object describing the internal structure of hadrons, the parton distribution function, already introduced. Most of this discussion can be extended to other more differential and complex diagnostics of the structure of hadrons, and we invite an interested reader to dedicated reviews~\cite{Boussarie:2023izj,Diehl:2003ny}.

\begin{figure}[h!]
    \centering
    \includegraphics[width=1\textwidth]{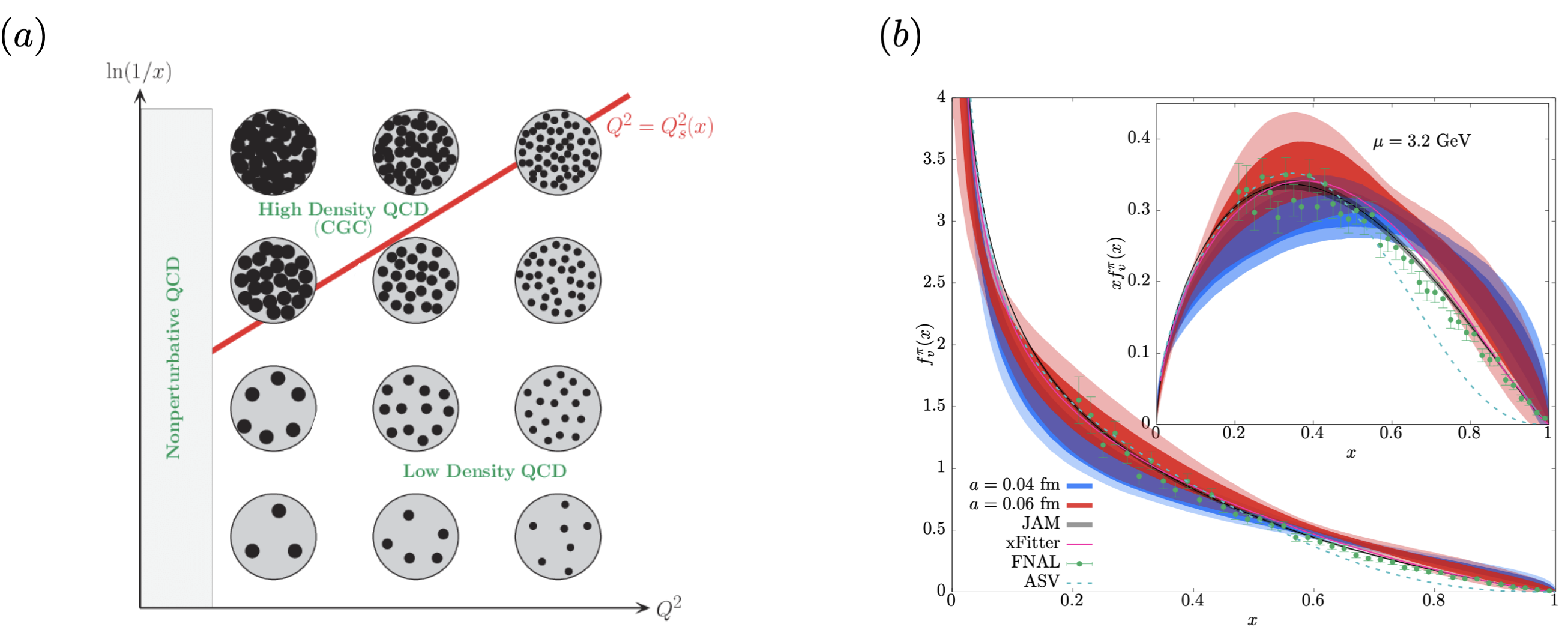}
    \caption{ Overview of PDF evolution and structure. $(a)$ Schematic QCD evolution in the \((\log 1/x,Q^2)\) plane, showing the different regimes for the internal properties of the hadron wavefunction. $(b)$ Valence pion PDF from lattice QCD compared with different phenomenological and Euclidean lattice extractions. Figures adapted from~\cite{Goncalves2015HeavyFlavourUltraHighCosmicRays, Gao:2020ito}, respectively.}
    \label{fig:pdf_LQCD}
\end{figure}

Historically, PDFs were first studied in the context of deeply inelastic scattering (DIS)~\cite{Bjorken:1969ja,Breidenbach:1969kd}. In these experiments, one shoots a highly boosted electron against a hadron, typically a proton or an ion. As the electron is highly boosted, it can resolve the internal structure of the target, giving information about the internal dynamics inside hadrons. These properties can be inferred from the final state particles, which consist of the scattered electron and the remnants of the hadron. Thus,
the properties of the scattering experiment are parametrized in terms of two Lorentz-invariant variables: the virtuality of the exchanged photon, $
Q^2 = -q^2$, and the Bjorken-\(x\), $
x = \frac{Q^2}{2P \cdot q}$, with \(q\) the momentum transferred from the electron to the hadron, and
\(P\) is the hadron momentum. The scale \(Q\) controls the resolution of the probe, while \(x\) is interpreted, in the parton model of QCD~\cite{Feynman:1969ej}, as the fraction of the hadron longitudinal momentum carried by the struck parton. As a result, PDFs $f_{i/h}$ give the density of partons of type \(i\) carrying momentum fraction \(x\) when probed at scale \(Q\), and similarly to the two to $n$ scattering example discussed above, they enter directly the factorization formula for the inclusive DIS cross-section~\cite{Collins:2011zzd}:
\begin{equation}
d\sigma_{eh\to eX}
=
\sum_{a=q,\bar q,g}
\int_0^1 dx\,
f_{a/h}(x,\mu_F)\,
d\hat{\sigma}_{ea\to e X}
\!\left(q,x P;\mu_F\right)
+
\mathcal{O}\!\left(\frac{\Lambda_{\rm QCD}^2}{Q^2}\right)\, .
\label{eq:DIS-factorization}
\end{equation}
 We note that the PDF appearing here is the same as in Eq.~\eqref{eq:factorization}, as an explicit example of universality. Again treating the partonic cross-section in perturbation theory, this process gives a clean channel to measure PDFs from data, as one needs only to compute the electron-parton scattering process perturbatively and then invert Eq.~\eqref{eq:DIS-factorization}. Such a program, in a much evolved form and combining input from several experimental set-ups, has been employed to obtain fits for the different PDFs of hadrons, forming the baseline for a large extent of our understanding of structure of hadrons at high energies~\cite{Gao:2017yyd,Hou:2019efy,Bailey:2020ooq,NNPDF:2021njg,Eskola:2021nhw}. 

 In Fig.~\ref{fig:pdf_LQCD} we summarize the basic current understanding of PDFs. In panel $(a)$, we show the evolution of the PDF in the $(\log 1/x, Q^2)$ plane, as obtained from the (perturbative) understanding in QCD. At fixed $x$, as $Q$ increases, meaning a larger momentum exchange between the probing electron and the hadron, one observes that the hadron is resolved
with finer spatial resolution, deviating from the valence picture of three quarks (for the proton) as indicated in the left bottom corner. In contrast, at fixed $Q$, the hadron becomes increasingly more saturated by an increasing number of partons carrying smaller $x$. In this regime the gluon density grows rapidly: when the density becomes sufficiently large, the hadron becomes saturated, indicated by the line $Q_s(x)$, and non-linear evolution effects become important to satisfy unitarity, see e.g.~\cite{Kovchegov:2012mbw}. This theoretical picture is reflected in panel $(b)$, which shows the valence quark PDF of the pion. The distribution is concentrated at intermediate \(x\), where the pion momentum is shared between its valence constituents, i.e. quark-antiquark pair. It is suppressed as \(x\to 1\), where a single valence quark would have to carry almost all of the pion momentum. This particular PDF is rather insensitive to the small-$x$ dynamics, as it couples only to the valence structure.

\begin{figure}[h!]
    \centering
    \includegraphics[width=1\textwidth]{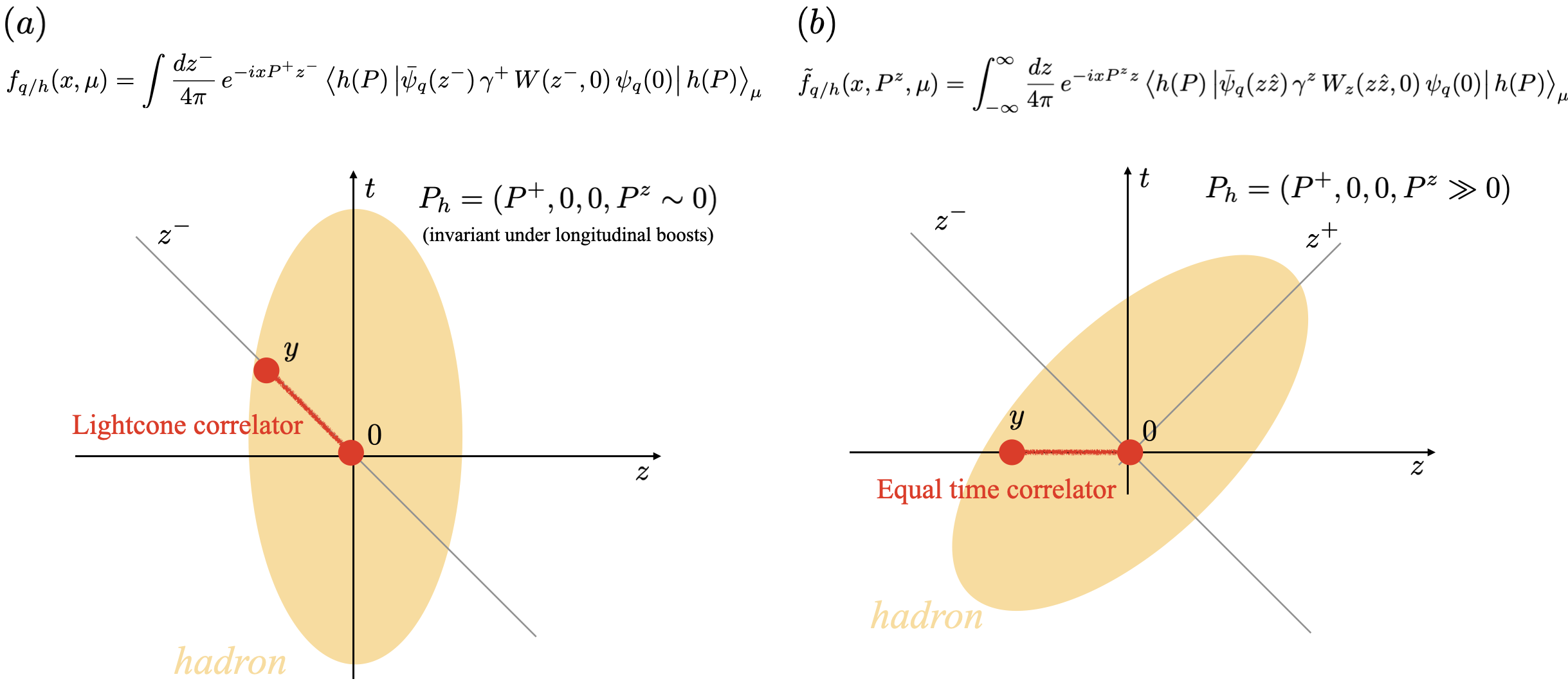}
    \caption{Comparison between lightcone and equal-time correlators for PDF computations. $(a)$ The standard PDF is defined from a lightcone quark correlator in a highly boosted hadron. $(b)$ In the quasi-PDF approach, this is replaced by a spatial correlator evaluated in a hadron state with large momentum \(P^z\), which can be related to the PDF through perturbative matching~\cite{Ji:2013dva,Ji:2014gla}.}
    \label{fig:pdf_fig}
\end{figure}

So far, the description of the PDF we have presented relies on perturbative understanding of quantum evolution in QCD, and the possibility to perform a data driven extraction, following the factorization formula in Eq.~\eqref{eq:DIS-factorization}. Complementary to this, and following a more detailed analysis of QCD factorization for DIS, one can directly write the PDF in terms of QCD operators~\cite{Collins:1981uw}:
\begin{equation}
f_{q/h}(x,\mu)
=
\int \frac{dz^-}{4\pi}\,
e^{-i x P^+ z^-}
\left\langle h(P)\left|
\bar{\psi}_q(z^-)\,\gamma^+\,
W(z^-,0)\,
\psi_q(0)
\right|h(P)\right\rangle_{\mu} \, .
\label{eq:pdf-definition}
\end{equation}
Here $|h(P)\rangle$ corresponds to the hadron state with momentum $P$, while $\psi_q(z)$ is the quark field at spacetime position $z$, $\gamma^+$ is the Dirac $+$-gamma matrix, and $W(z^-,0)$ is a gauge link connecting the origin to a lightlike-separated spacetime point with $z^+=0$ and zero transverse separation. We use $z^\pm=(t\pm z)/\sqrt{2}$, $P^\pm=(P^0\pm P^z)/\sqrt{2}$, and $\gamma^\pm=(\gamma^0\pm\gamma^3)/\sqrt{2}$. Heuristically, the  operator form of the PDF is natural, as the operator $\bar \psi \gamma^+\psi$ can be understood as the quark number operator on the lightcone. This structure is further illustrated in Fig.~\ref{fig:pdf_fig} $(a)$.

Equation~\eqref{eq:pdf-definition} makes clear the conceptual difficulties in extracting PDFs using non-perturbative methods. The first challenge lies in the preparation of the target hadron state $|{h(P)}\rangle$, due to complex particle spectrum of QCD. This aspect has been, in part, addressed in Euclidean QCD, where one typically constructs interpolating operators with
the quantum numbers of the desired hadron state, and extracts their (Euclidean) time correlation functions. As these correlators admit a spectral decomposition
with contributions weighted by \(e^{-E_n \tau}\), large Euclidean-time $\tau$ separations
exponentially suppress excited states, allowing the lowest-energy state in the chosen quantum-number channel to be isolated, see e.g.~\cite{GattringerLang2010,Luscher:1990ck}.

The second problem is, in some sense, more fundamental. The operator in
Eq.~\eqref{eq:pdf-definition} is defined at a light-like separation. As a result, a direct extraction of PDFs, and related hadron structure diagnostics, is not possible using standard lattice QCD if one follows the formulation in Fig.~\ref{fig:pdf_fig} $(a)$. Remarkably, this difficulty in accessing light-like correlators can be \textit{avoided} following a non-trivial connection between the standard PDF and the spatial correlation function shown in panel $(b)$. Following the original proposal by Ji~\cite{Ji:2013dva}, one can \textit{trade} the unequal-time correlator, by a purely spatial separation, but now measured in a proton state carrying a large momentum $P^z$. This gives rise to a so called quasi-PDF $\tilde  f_{q/h}$, in general distinct from standard PDF. However, at very large momentum $P^z$, when the hadron momentum approaches the lightcone, one can show that these two non-perturbative objects match in the infrared, while differing at short distances. Thus, by computing the quasi-PDF, one can extract the correct low energy form of the PDF, while ultraviolet behavior can be treated using perturbative matching to the PDF~\cite{Ji:2014gla,Ji:2020ect}. Thus, and in the spirit of the discussion of factorization in QCD, these two objects can be schematically related as:
\begin{equation}
\tilde f_{q/h}(x,P^z,\mu)
=
\int_{-1}^{1}\frac{dy}{|y|}\,
C\!\left(\frac{x}{y},\frac{\mu}{|y|P^z}\right)
f_{q/h}(y,\mu)
+
\mathcal{O}\!\left(
\frac{\Lambda_{\rm QCD}^2}{(P^z)^2},
\frac{M_h^2}{(P^z)^2}
\right) \, ,
\end{equation}
where the matching kernel $C$ provides the ultraviolet correspondence connecting these two distribution functions, which can be computed perturbatively, with $M_h$ the hadron mass. Although here we discussed the particular case of hadron PDFs, one should note that this procedure can be applied to more complex objects, such as transverse momentum distribution functions, generalized PDFs, and the connection between the Euclidean and Minkowski correlators has been systematized under the umbrella of Large-Momentum Effective Theory~\cite{Ji:2014gla,Ji:2020ect,Boussarie:2023izj}. 
We also note that this construction is not unique, see e.g.~\cite{Radyushkin:2017cyf}, although the key idea of relating light-like and spatial correlators is generally employed.

Returning to Fig.~\ref{fig:pdf_LQCD} $(b)$ in addition to the extractions obtained from data driven approaches (data points), two curves obtained at different lattice spacing $a$ are also presented, following the quasi-PDF construction, see also~\cite{Orginos:2017kos,Radyushkin:2019owq} for other lattice studies. One observes that the lattice approach has a good agreement with the other extractions, in the support of the data points, especially for intermediate values of $x$. Nonetheless, this agreement becomes worse as $x\to 1$, where statistical limitations, finite-momentum effects, and enhanced power corrections become important. At the other extreme, small-$x$ extractions are particularly challenging because of finite hadron momentum, the accessible range of spatial separations, and the lack of QCD resummation in $\log \frac{1}{x}$. Current Euclidean lattice extractions are therefore most reliable at intermediate $x$ values.\footnote{Note that the valence PDF in Fig.~\ref{fig:pdf_LQCD} $(b)$ does not directly probe the gluon-dominated saturation dynamics shown in panel $(a)$.}

In summary, the present description of hadrons at high energies relies on two complementary strategies. Data driven extractions use QCD factorization, universality, and perturbative evolution to infer PDFs from scattering data, while lattice approaches provide a first-principles route by relating lightcone PDFs to spatial correlation functions evaluated in boosted hadron states. Despite their success, both approaches have limitations: global fits are restricted by the available data and by the assumed factorization framework, whereas lattice extractions are currently most reliable at intermediate \(x\) and face systematic challenges from finite hadron momentum, and perturbative matching. This leaves open important regions of hadronic structure to be explored, and strategies that can directly probe Eq.~\eqref{eq:pdf-definition} from first principles in QCD would be of extreme value to map the internal structure of hadronic matter.

\subsection{Emergent QCD matter in ultrarelativistic  collisions}\label{subsec:HICs}
In Section~\ref{subsec:pp} we discussed the framework commonly used to describe high energy scattering in QCD. There the focus was on processes characterized by a hard momentum scale $Q$, such as a large momentum transfer, which allows the dynamics to be factorized into distinct contributions associated with well-separated scales. This is the setting in which perturbative QCD, supplemented by non-perturbative but universal objects such as PDFs, provides a controlled description of collider observables.

However, many high energy nuclear collisions are not dominated by a single hard scale. Instead, a substantial fraction of the available energy is converted into the production of a large number of relatively soft hadrons. At first sight, such events may appear less tractable, since they do not come with an obvious perturbative scale on which to organize the calculation. Yet precisely for this reason they can probe a different and particularly rich aspect of QCD: the emergence of many-body dynamics from strongly interacting degrees of freedom.

Ultrarelativistic heavy-ion collisions (HICs) provide the cleanest experimental realization of this situation. In these collisions, two highly boosted nuclei overlap for a very short time and deposit a large amount of energy over a region whose transverse size is set by the nuclear geometry. Compared to $pp$ collisions, the large number of participating nucleons deposits a high energy density in a small spacetime region, resulting in a substantially larger production of particles in a small spatial patch. The matter produced immediately after the collision is therefore a highly excited, far-from-equilibrium QCD state, which makes HICs a principal experimental setting for exploring such states in the theory. After this state is produced, the rapid interaction rate between particles in the bulk matter leads to an accelerated expansion and equilibration of the deconfined matter. At this stage, when the bulk medium is near equilibrium, its dynamics can be effectively described by nearly ideal relativistic hydrodynamics. The onset of hydrodynamic behavior need not coincide with complete local thermal equilibrium, and it further drives the expansion of the medium, controlled by temperature gradients. This results in the collective flow of matter. As the medium cools through the QCD crossover region, confinement becomes effective again: quarks and gluons are converted into hadrons, and these hadronic remnants carry the imprint of the preceding collective expansion to the experimental detectors, see~\cite{Busza:2018rrf,Berges:2020fwq} for reviews.

A key signature of this collective behavior is the azimuthal anisotropy of the final-state particle distribution. It is conventionally characterized by the flow (Fourier) coefficients $v_n$, defined through~\cite{Ollitrault1992,VoloshinZhang1996,PoskanzerVoloshin1998,HeinzSnellings2013}
\begin{equation}
\frac{dN}{d\phi}
=
\frac{N}{2\pi}
\left[
1+2\sum_{n=1}^{\infty}
v_n \cos n(\phi-\Psi_n)
\right],
\end{equation}
where $\phi$ is the azimuthal angle of the detected hadron and $\Psi_n$ denotes the orientation of the corresponding harmonic plane, roughly set by the geometry of the colliding nuclei. These observables are useful because they characterize how the initial geometry of the collision is converted into the final momentum distribution of the produced hadrons. For example, the most studied coefficient, $v_2$ (elliptic flow), is expected to deviate significantly from zero for off-central collisions, where the partially overlapping, almond-like shape of the incoming nuclei should lead to a sizeable flow signal due to geometry-induced pressure gradients in the bulk. The higher harmonics are usually associated with fluctuations in the incoming wavefunctions, being suppressed as $n$ increases. Thus, the $v_n$'s indicate an inversion of the initial deformation in coordinate space, which is progressively washed out, reappearing as an anisotropy in momentum space. This response is often summarized schematically as $
v_n \simeq \kappa_n \epsilon_n$, where $\epsilon_n$ is the $n$th harmonic eccentricity of the initial energy-density profile and $\kappa_n$ encodes the response of the medium. The fact that final-state hadrons retain this information about the initial geometry provides direct evidence that they are not emitted independently, but instead originate from a collectively evolving many-body state.

\begin{figure}[h!]
    \centering
    \includegraphics[width=1\textwidth]{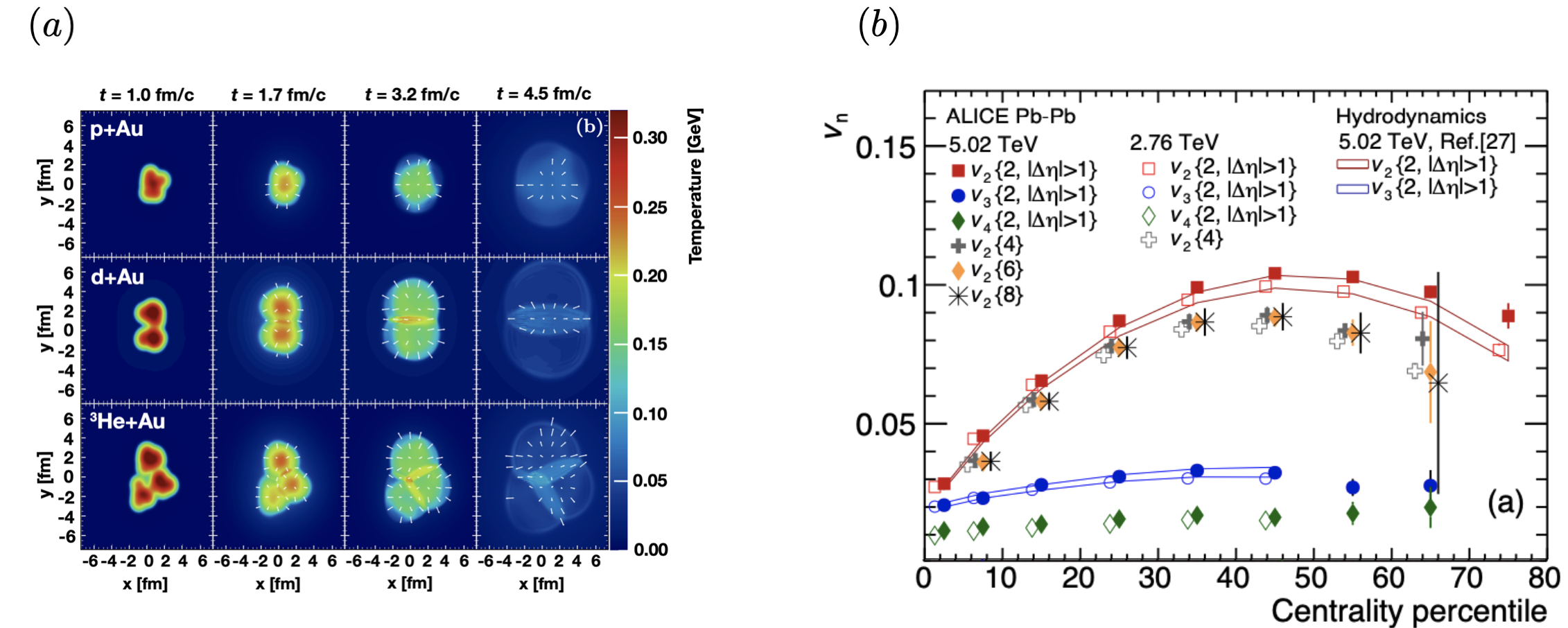}
    \caption{Emergence of collective flow in high energy nuclear collisions. $(a)$ Hydrodynamic evolution of the transverse temperature profile for different initial projectile geometries in small collision systems. The initially localized and anisotropic energy density profiles rapidly expand and become smoother, while their geometric deformation is converted into collective motion. $(b)$ Measurements of the anisotropic-flow coefficients $v_n$ in lead-lead collisions compared with hydrodynamic calculations. The non-zero values of $v_2$, $v_3$, and higher harmonics show that the produced particles are not emitted independently, but retain a collective memory of the initial geometry and its event-by-event fluctuations. Figures adapted from~\cite{PHENIX:2018lia,ALICE:2016ccg}, respectively.}
    \label{fig:collectivity_HIC}
\end{figure}

This mechanism is illustrated in Fig.~\ref{fig:collectivity_HIC}. Panel $(a)$ shows hydrodynamic simulations of the transverse temperature profile for different initial geometries, for small colliding systems. The initially deformed profiles expand and become smoother with time, while their geometric information is transferred into the collective velocity field. Panel $(b)$ shows the measured anisotropic-flow coefficients in lead--lead collisions, together with hydrodynamic theoretical calculations. The observation of sizable non-zero $v_n$ coefficients, and their agreement with hydrodynamic modeling, supports the interpretation that the soft particles produced in heavy-ion collisions originate from a collectively expanding QGP rather than from independent particle emission. Such observations have enabled precise extractions of flow coefficients and, through comparison with viscous hydrodynamic simulations, of the transport properties of QCD matter. This has resulted in the remarkable discovery that the QGP produced in HICs behaves as an almost ideal fluid, with an exceptionally small viscosity~\cite{HeinzSnellings2013}.

Despite this remarkable progress, several conceptual questions remain open. Panel~$(a)$ highlights one of them: signatures of collective behavior appear to persist even in small collision systems, where the system size and lifetime are much shorter, and the gradients much larger, than in conventional heavy-ion collisions. This challenges our understanding of the applicability window of hydrodynamics, which is expected to describe systems dominated by long-wavelength modes and sufficiently small gradients. Thus, although hydrodynamic calculations have achieved remarkable phenomenological success, understanding how this regime emerges from the underlying QFT dynamics, and when it ultimately breaks down, remains an unresolved problem.

Moreover, even when hydrodynamics successfully describes the bulk evolution, the theoretical connection between this near-equilibrium stage and the preceding far-from-equilibrium QCD dynamics remains incomplete. Since the observed hadrons mostly carry information about the later collective evolution of the medium, they provide only indirect access to the earliest stages of the collision. Probing how the initial many-body state forms and approaches the hydrodynamic regime therefore requires new theoretical approaches capable of addressing real-time, out-of-equilibrium QCD dynamics.

Beyond the characterization of the bulk medium through soft collective flow, one can also probe the QGP by studying the evolution of energetic or massive states produced on top of it. Such \textit{hard probes} can be of interest to, for example, address the properties of matter beyond hydrodynamics, as they do not thermalize jointly with the bulk matter, but rather evolve in parallel to it~\cite{Apolinario:2022vzg,Cao:2020wlm}. These probes are thus found in rarer HICs events, where in addition to the produced bulk matter, a hard interaction takes place at early times that generates a high energy or massive state. Since this hard process involves a large momentum scale, its production can be described using the factorization picture discussed in Section~\ref{subsec:pp}. The subsequent propagation of the produced energetic partons through the QGP, however, is sensitive to the properties of the medium. In what follows we discuss two examples of hard probes.

The first are jets, which originate from energetic quarks and gluons produced in an early hard scattering. These partons subsequently undergo a cascade of QCD radiation, forming collimated sprays of particles; see Fig.~\ref{fig:pp_fig} $(a)$ where a multi-jet event is shown. In heavy-ion collisions, these particle showers evolve while traversing the QGP, exchanging energy and momentum with the surrounding medium. The resulting phenomenon, known as jet quenching~\cite{Bjorken:1982tu,Wang:1992qdg}, leads to modifications of the observed jet spectrum and substructure relative to the corresponding $pp$ baseline, allowing to infer the properties of the underlying medium.

\begin{figure}[h!]
    \centering
    \includegraphics[width=1\textwidth]{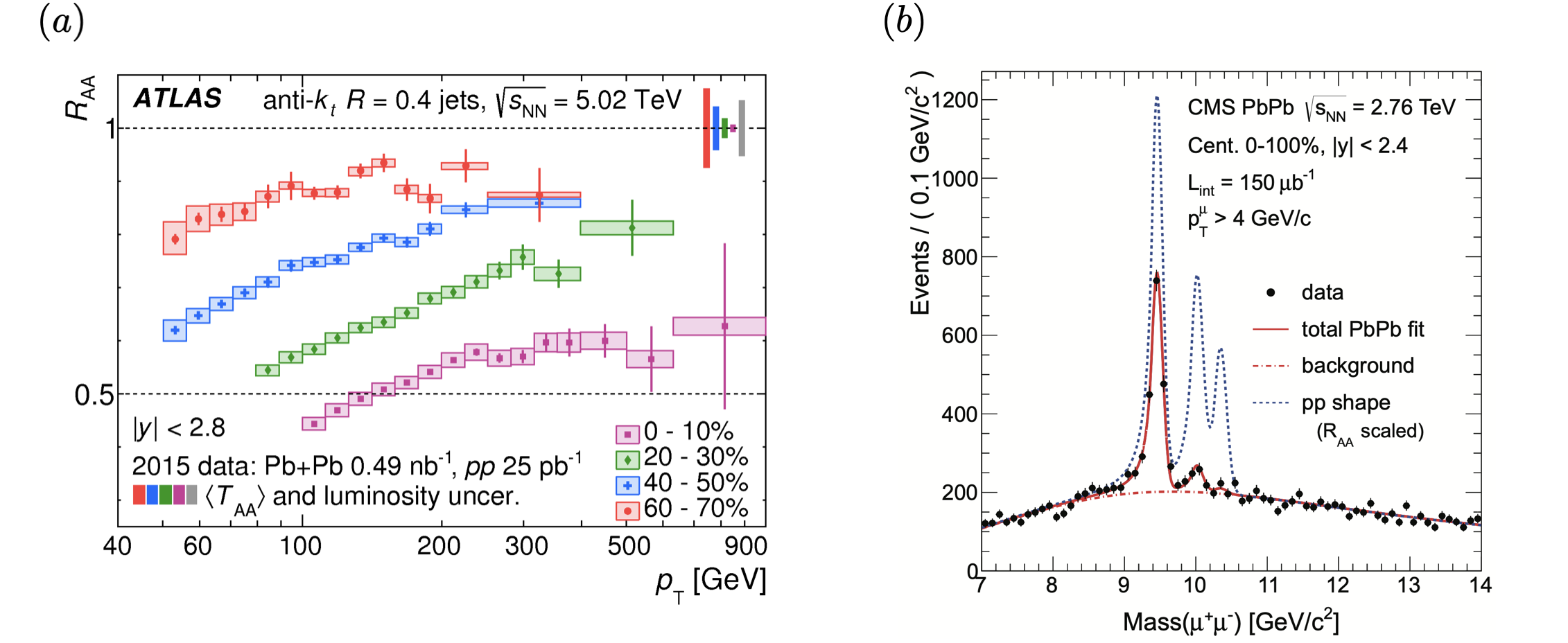}
    \caption{Hard probes of the QGP.
$(a)$ Nuclear suppression factor of high momentum jets, reflecting parton energy loss in the medium.
$(b)$ Suppression of quarkonium resonances, illustrating how the QGP dissociates heavy-quark bound states according to their binding energies. Figures adapted from~\cite{ATLAS:2018gwx,CMS:2016rpc} and~\cite{Satz:2013xja,CMS:2012gvv}, respectively.}
    \label{fig:jetsquarkonia_HIC}
\end{figure}

This effect is illustrated in Fig.~\ref{fig:jetsquarkonia_HIC} $(a)$ through the nuclear modification factor $R_{AA}$. This observable compares the number of jets in HICs to that in proton-proton collisions, rescaled by the nuclear overlap. Values $R_{AA}<1$ indicate suppression relative to the scaled $pp$ baseline; at high transverse momentum, final-state energy loss is a dominant contribution, including medium-induced radiation. Schematically, this suppression can be encoded in a quenching factor $\textbf{Q} = \exp \left\{- \int d\omega \int d^2k \, \frac{dI}{d\omega d^2k}  \left(1- e^{-\frac{\nu\omega}{p_T}}\right)\right\}$, 
where $dI/(d\omega d^2k)$ denotes the medium induced gluon radiation spectrum, $p_T$ is the jet transverse momentum, $\nu$ is the dimensionless spectral-slope parameter entering this approximation, and $R_{AA}\propto \textbf{Q}$, see e.g.~\cite{Landau:1944if,Baier:2001yt}. Thus, the observed suppression in panel $(a)$ constrains properties of the radiative energy loss dynamics in the bulk, and provides indirect access to real-time properties of the QGP. Although the radiative spectrum, and other quantities alike can be treated using perturbative and lattice methods, they rely on several simplifying assumptions, which typically reduce the complexity of the bulk matter description~\cite{Apolinario:2022vzg}. Thus, important progress in understanding how jets can be used to map HICs events requires new approaches which allow to combine partonic cascades with the real-time quantum evolution of the bulk. 

A second important class of hard probes is provided by heavy-quark bound states, i.e. quarkonia. Since the heavy quarks are mainly produced in the earliest hard scatterings, due to their large mass, their production rate can be constrained from the corresponding $pp$ baseline. Moreover, and in parallel to jets, they do not fully thermalize with the medium, and instead propagate through the QGP as external probes. The presence of the medium reflects itself by the modification of the effective interaction between the heavy quarks, and was originally proposed~\cite{Matsui:1986dk} to lead to the Debye screening of their confining potential. This results in a weakening of the respective binding  energies of the quarkonium states, eventually leading to their dissociation in the medium.

This simple picture naturally drives the expectation of a sequential suppression of quarkonium states, ordered by their binding energies. This is illustrated in Fig.~\ref{fig:jetsquarkonia_HIC}~$(b)$, where the dimuon invariant-mass spectrum shows the bottomonium resonances $\Upsilon(1S)$, $\Upsilon(2S)$, and $\Upsilon(3S)$ (increasingly smaller binding energy states). In the $pp$ baseline the three resonance peaks are clearly visible, while in HICs the excited states are strongly reduced relative to the ground state. This ordered suppression is consistent with the expectation that the QGP modifies the in-medium heavy-quark potential and can dissociate quarkonium states according to their binding energy.

While this qualitative interpretation is consistent with the original Debye-screening picture, the full dynamics of quarkonium production in heavy-ion collisions, including dissociation, regeneration, and feed-down, is more involved, as has been widely explored over the last decades~\cite{Andronic:2024oxz,Akamatsu:2020ypb,Rothkopf:2019ipj,Brambilla:2016wgg}. A first-principles understanding of quarkonium suppression from QFT, including the real-time evolution of heavy-quark bound states inside the bulk medium, would therefore provide a more direct connection between the observed yields and the microscopic properties of the QGP.

\section{The Quantum Information Science toolkit}\label{sec:QIS}
Having identified several open problems in high energy nuclear physics where traditional QCD methods have limitations, we now turn the discussion to developments in QIS that may provide complementary ways forward. We first review a set of computational and simulation techniques designed to study the properties and dynamics of quantum many-body systems. We then explain how these methods can be applied to lattice gauge theories, using the Schwinger model as a concrete example.

\subsection{Quantum inspired classical methods}
Let us suppose that we want to study the real-time properties of a quantum mechanical system characterized by a Hamiltonian $H$ acting on a Hilbert space $\mathcal{H}$. For simplicity, we consider a system made of $N$ spin-$1/2$ degrees of freedom, although the discussion generalizes to other scenarios. In this case the Hilbert space, $\mathcal{H} = \bigotimes_{i=1}^{N} \mathbb{C}^2$, becomes finite dimensional, ${\rm dim}\,\mathcal{H}=2^N $. Any Hamiltonian acting on this Hilbert space can be decomposed in a basis of Pauli strings,
\begin{equation}\label{eqPaulis}
    H = \sum_{\boldsymbol{\alpha}} c_{\boldsymbol{\alpha}} P_{\boldsymbol{\alpha}},
    \qquad
    P_{\boldsymbol{\alpha}}
    =
    \bigotimes_{i=1}^{N} \sigma_i^{\alpha_i},
    \qquad
    \alpha_i \in \{0,x,y,z\},
\end{equation}
where $\sigma^0_i\equiv 1_i$, $\sigma_i^{x,y,z}$ are Pauli matrices acting on site $i$, and $c_{\boldsymbol{\alpha}}$ specify the model's Hamiltonian. In this language, the real-time evolution of an initial state $|\psi(0)\rangle$ is given by
\begin{equation}\label{eq:ooo}
       |\psi(t)\rangle  = e^{-iHt} |\psi(0)\rangle ,
\end{equation}
and observables are obtained from expectation values, e.g.
$ \langle \mathcal{O}(t) \rangle =\langle \psi(0)| e^{iHt} \mathcal{O} e^{-iHt} |\psi(0)\rangle$.

The most direct (classical) method to solve Eq.~\eqref{eq:ooo} is to find the similarity matrix which diagonalizes $H$. Extracting the energy eigenvectors $|n\rangle$ and eigenvalues $E_n$, the real-time dynamics become trivial as
\begin{align}
        |\psi(t)\rangle = \sum_n c_n^0 e^{-i E_n t} |n \rangle\, ,
\end{align}
where $c_n^0$ defines the initial state in the energy basis. This strategy, known as exact diagonalization (ED), should be familiar from elementary quantum mechanics, as it is the standard approach for exactly solving small quantum systems, such as models made of a few spins. Within its range of applicability, ED is extremely useful: it provides exact numerical access to the theory's spectrum, and therefore serves as an essential benchmark for other approximate methods which can be used for larger system sizes. However, for many-body systems this approach quickly becomes impractical, as even storing a generic state in memory would be exponentially costly. For example, in double precision, a state vector requires $16\,2^N$ bytes, since each complex amplitude contains two 64-bit floating-point numbers. Thus, for $N=30$ one would need about $16$ GB of memory, while for $N=32$ the number increases to $64$ GB to store a single state. If the Hamiltonian is dense, storing it in memory would require approximately $16\,2^{2N}$ bytes, so that already for $N=16$ this would imply a memory storage of $64$ GB, even before diagonalization. Consequently, full ED is typically limited to $N\lesssim \mathcal{O}(20)$ spins.

\begin{figure}[h!]
    \centering
    \includegraphics[width=1\textwidth]{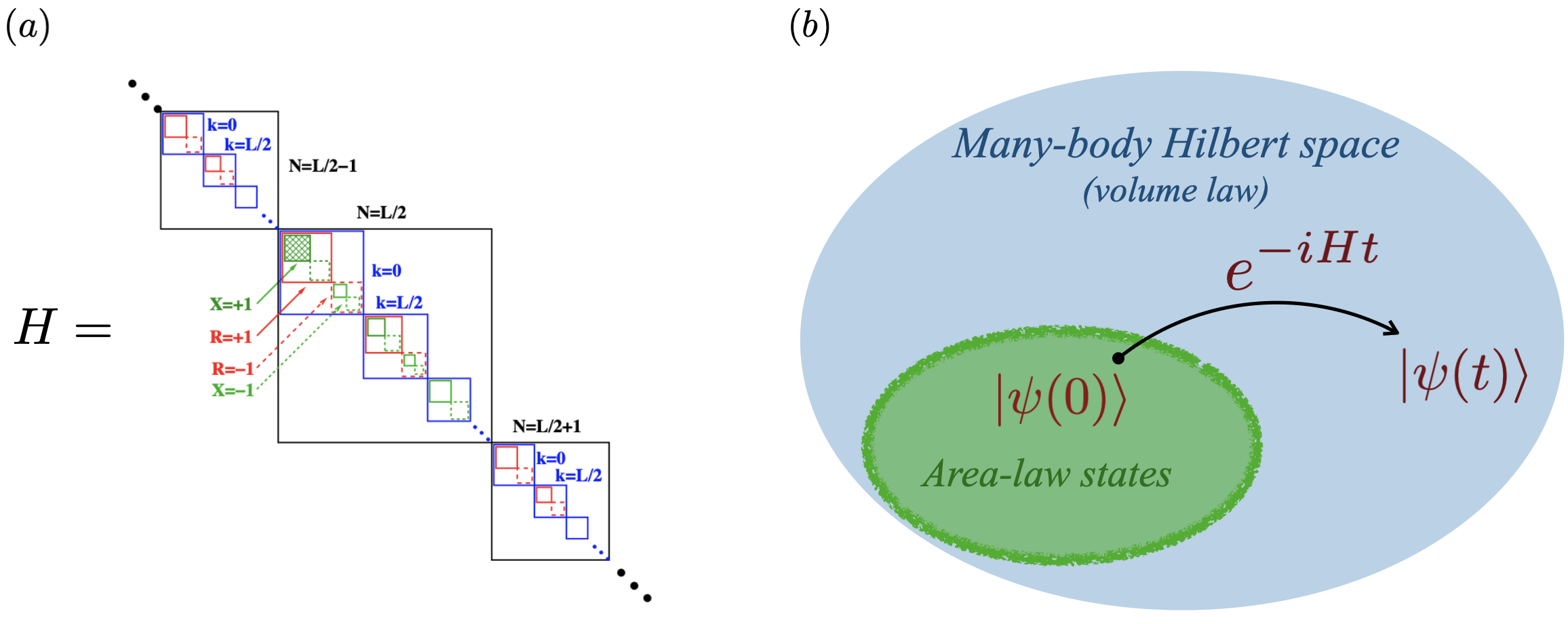}
    \caption{Classical strategies for many-body Hamiltonians.
$(a)$ Locality and symmetries often make the Hamiltonian sparse and block diagonal in a suitable basis, allowing ED and Krylov-based methods to access larger systems than a naive dense diagonalization approach. Figure adapted from~\cite{Jung:2020ouf}.
$(b)$ Generic many-body states occupy an exponentially large Hilbert space and can develop volume-law entanglement under time evolution, while tensor network methods target the smaller manifold of low entanglement, area-law states.}
    \label{fig:ED+TN}
\end{figure}

The accessible system size can be extended when the Hamiltonian has additional structure. This is often the case in physical systems: local interactions make the Hamiltonian sparse in a simple product basis, since each basis state is connected only to a limited number of other states. Symmetries further decompose the Hilbert space into independent sectors labeled by conserved quantum numbers, so that one only needs to diagonalize smaller blocks of $H$. Even in such cases, typically $N\lesssim \mathcal{O}(40)$, when locality and symmetries are efficiently exploited. This structure is illustrated in Fig.~\ref{fig:ED+TN} $(a)$, where the Hamiltonian has been block diagonalized into sectors of different quantum numbers $N$ and $k$. On top of this, ED approaches can be made more powerful when coupled to approximate methods which do not fully diagonalize $H$. Typical examples are Lanczos and Arnoldi algorithms~\cite{Lanczos1950,Arnoldi1951,Saad1992,Lehoucq1998}, where one constructs a (Krylov) subspace generated by repeated applications of the Hamiltonian to a given initial vector state:
\begin{equation}
    \mathcal{K}_m(H,|\psi\rangle)
    =
    {\rm span}
    \left\{
    |\psi\rangle,
    H|\psi\rangle,
    H^2|\psi\rangle,
    \dots,
    H^{m-1}|\psi\rangle
    \right\}.
\end{equation}
Then, one can solve for the dynamics and properties of the model in this subspace. As long as most of the relevant evolution of the system's wavefunction is localized in this sector, then the description should be close to the exact ED solution, at a much smaller numerical cost. These and other methods significantly extend the practical reach of ED for sparse Hamiltonians, and we refer the reader to dedicated reviews for further details~\cite{Lanczos1950,ParkLight1986,Saad1992,HochbruckLubich1997,Lehoucq1998,Sandvik2010,Dagotto1994}. 

The key point/limitation for ED methods is that they inherit the exponential growth of the many-body Hilbert space, and as a result their applicability to the study of, e.g., quantum field theories is conceptually limited, as these in the continuum have a large infinite dimensional Hilbert space. This limitation motivates the development of other methods that allow to explore the many-body dynamics in quantum systems. Such approaches generally approximate the relevant states or dynamics without explicitly exploring the full Hilbert space.

The construction of approximate methods to represent quantum dynamics is not unique, and here we discuss the physical reasoning behind the tensor network (TN) approach~\cite{Fannes1992,White1992,Vidal2003,VerstraeteCirac2004}. As illustrated schematically in Fig.~\ref{fig:ED+TN} $(b)$, in general, a many-body Hilbert space contains qualitatively different classes of states. Generic states occupy the largest part of the Hilbert space and typically exhibit volume-law entanglement, meaning that the entanglement entropy of a subsystem scales with its volume. Such states are genuinely many-body quantum objects and require exponentially many parameters to describe their wavefunction fully. On the other hand, many physically relevant low energy states of local gapped Hamiltonians satisfy an area-law~\cite{Hastings2007,EisertCramerPlenio2010}: their entanglement scales only with the size of the boundary separating a subsystem from its complement. 

This observation, i.e. that for gapped local theories one expects an area-law sector, is particularly powerful in one spatial dimension, as partitioning a system creates only a zero-dimensional boundary, so area law states have entanglement that remains bounded as the system size increases. This suggests that many relevant one dimensional states can be described without exploring the full exponentially large Hilbert space. Tensor network methods make this idea concrete, by allowing to design efficient representations of quantum states which are truncated based on their entanglement. Thus, if the targeted state satisfies area-law, then a TN representation with a small truncation should (nearly) exactly capture the form of physical state. The driving principle behind this construction is thus to represent a complex many-body wavefunction as tensor product of simpler tensors, with a control on the entanglement between them.

To illustrate this, consider first a completely unentangled product state of $N$ spins,
\begin{equation}
    |\psi_{\rm prod}\rangle
    =
    |\psi_1\rangle \otimes |\psi_2\rangle \otimes \cdots \otimes |\psi_N\rangle \, ,
\end{equation}
where each local state can be written as $|\psi_i\rangle= a_i |\uparrow\rangle_i + b_i |\downarrow\rangle_i $. Such a state is specified by the local coefficients $(a_i,b_i)$ at each site, and therefore requires only a number of parameters that grows linearly with $N$. It is much simpler than a generic state in the theory:
\begin{equation}\label{eq:genral_qstate}
    |\psi\rangle
    =
    \sum_{s_1,\ldots,s_N}
    c_{s_1\cdots s_N}
    |s_1\cdots s_N\rangle \, ,
    \qquad
    s_i=\uparrow,\downarrow \, ,
\end{equation}
which requires specifying $\mathcal{O}(2^N)$ coefficients, as discussed above. Tensor network methods offer a way to interpolate between the trivial product state, which is (nearly) classical, and a more general (low-entanglement) state. This is best illustrated via the matrix product state (MPS) construction~\cite{Schollwock:2011, Cirac:2020obd}, which is one particular class of TNs. In essence, MPS adopt the product state form but allow for neighboring sites to be connected by auxiliary indices that carry entanglement. More explicitly, the MPS proposes that an arbitrary state, as in Eq.~\eqref{eq:genral_qstate} can be (approximately) represented by:
\begin{equation}\label{eq:mps_ansatz}
    |\psi_{\rm MPS}\rangle
    =
    \sum_{s_1,\ldots,s_N}
    A^{s_1}_1 A^{s_2}_2 \cdots A^{s_N}_N
    |s_1 s_2 \cdots s_N\rangle \, ,
\end{equation}
where $s_i$ labels a basis of the local Hilbert space at site $i$, and the $A_i^{s_i}$ are matrices whose auxiliary indices are contracted along the chain. The internal dimension of these auxiliary indices is called the bond dimension $\chi$. For $\chi=1$, the MPS reduces to a product state, and increasing $\chi$ allows Eq.~\eqref{eq:mps_ansatz} to represent increasingly more entangled states. For a finite chain, any state can be represented exactly with sufficiently large bond dimension. Area-law states often admit accurate representations with modest $\chi$, while volume-law states generally require a bond dimension that grows exponentially with system size. Finally, the bond dimension can be directly connected to the entanglement growth from the Schmidt decomposition across a bipartition of the chain. Since an MPS has at most $\chi$ non-zero Schmidt coefficients across any cut, the entanglement entropy satisfies $S \leq \log \chi$. As a result, the MPS construction directly limits the amount of entanglement retained across each cut.

Up to this point, we have only solved part of the problem, namely the efficient representation of certain quantum states. However, if one is interested in real-time dynamics, one must also be able to efficiently prepare low-energy states of the theory and act on them with, e.g., the unitary time evolution operator. Note that both these steps are necessary to realize Eq.~\eqref{eq:ooo}. It turns out that both these tasks can be implemented efficiently within the TN framework. Originally motivated by the Wilsonian renormalization group idea~\cite{Wilson:1974mb,Wilson:1973jj}, the preparation of lowly entangled states from product states can be performed using the so called
density matrix renormalization group (DMRG) algorithm~\cite{White1992,Schollwoeck2005,Schollwoeck2011}. In spirit, this algorithm locally optimizes the local parameters of each MPS block such that the global ground state can be found. In its modern formulation, it can be understood as a variational optimization over the MPS manifold, where the energy functional 
\begin{equation}
    E[A]
    =
    \frac{
    \sum_i
    \langle \psi_{\rm MPS}[A]|h_{i,i+1}|\psi_{\rm MPS}[A]\rangle
    }{
    \langle \psi_{\rm MPS}[A]|\psi_{\rm MPS}[A]\rangle
    } \, .
\end{equation}
is minimized, with the $A$ the collection of MPS tensors and $H=\sum_i h_{i,i+1}$. In practice, this high dimensional optimization is made efficient by sweeping along the chain: all tensors are kept fixed except for one, and its parameters are locally optimized to find a lower energy state. This approach is particularly efficient for local Hamiltonians, although longer-range interactions can also be treated. The optimized tensor is then decomposed using a singular-value decomposition, and the sweep proceeds to the next site. Technical details on the DMRG algorithm can be found in~\cite{Orus:2014,Schollwock:2011,Schollwoeck2011}. Remarkably, although DMRG was originally developed as a ground-state variational algorithm, its main ingredients now underlie much of the modern TN toolkit. The same principles can be used for implementing dynamics, where after a short time evolution the obtained state is iteratively projected back to the low-entanglement subspace. This procedure can be done in a variety of ways, and we refer an interested reader to the relevant works~\cite{Vidal2004,WhiteFeiguin2004,DaleyKollathSchollwoeckVidal2004,Haegeman2016}.

Finally, one might wonder when does the TN approach fail. As discussed above, TNs extend the reach of ED by restricting the dynamics to the low-entanglement part of the Hilbert space, and are therefore optimally suited to one-dimensional systems whose relevant states satisfy an area-law.\footnote{We note, however, that TN approaches have also been extended to higher-dimensional systems, see e.g.~\cite{VerstraeteCirac2004,VerstraeteMurgCirac2008,Orus:2014}.} This restriction becomes a limitation for generic real-time evolution. Even if the initial state lies in the area-law region, as represented schematically in Fig.~\ref{fig:ED+TN} $(b)$, evolution under an interacting Hamiltonian generally produces entanglement across spatial bipartitions. This growth is often approximately linear at intermediate times, i.e. $ S(t) \sim v_E t$, where $v_E$ is an entanglement velocity~\cite{CalabreseCardy2005,CalabreseCardy2007,KimHuse2013}. Thus, as time increases, the state is driven away from the area-law manifold and towards the volume-law region of the many-body Hilbert space, as shown in Fig.~\ref{fig:ED+TN} $(b)$. Since an MPS with bond dimension $\chi$ can capture at most $S\lesssim \log \chi$ entanglement across a cut, faithfully following such dynamics requires a bond dimension that grows as $\chi \sim e^{S(t)} \sim e^{v_E t}$. Generic entanglement growth therefore restricts the times accessible to a fixed bond dimension MPS simulation.

Broadly, there are two possible ways to go beyond this limitation. The first is to consider alternative classical truncation schemes that are not based only on bipartite entanglement. For example, one may organize the simulation according to other measures of non-classicality, such as magic~\cite{Veitch2014,BravyiSmithSmolin2016,BravyiGosset2016}, or focus directly on the information needed to reconstruct local observables, as in local-information approaches~\cite{KvorningHerviouBardarson2022,ArtiacoLITE2024}. These methods can access classes of states that are difficult for standard TNs, such as highly entangled states with low magic or states whose long-range correlations are not important for local observables. Nevertheless, they still rely on identifying a restricted classical representation of the quantum state, and therefore can not efficiently describe arbitrary highly correlated quantum dynamics. See~\cite{Rudolph:2025gyq} for a recent example of an alternative numerical method.

The second route is to represent the dynamics of the target theory directly in a controllable device that itself obeys the laws of quantum mechanics. In such a device, highly entangled states are not stored as exponentially large classical arrays, but are realized as physical states of a quantum system. Thus, the obstruction is no longer rooted in the classical memory cost of representing the wavefunction. This is the perspective we turn to next.

\subsection{Quantum simulators, computers, and hybrid approaches}
The basic idea behind quantum simulation is to use a controllable quantum device to reproduce the states and dynamics of a target physical system. This can be achieved by engineering suitable interactions in the device or by implementing the target evolution through a sequence of quantum operations. The many-body state is then encoded directly in the physical degrees of freedom of the simulator, allowing highly entangled states to be represented without explicitly storing their exponentially many amplitudes in a classical memory.

There are several ways to realize this program, each with distinct advantages and limitations in terms of control, accuracy, and scalability. Different physical platforms offer different capabilities, and the most suitable implementation depends on the target system and the observables of interest. Here, we organize the discussion around three broad, partly overlapping strategies: analog quantum simulation, digital quantum simulation, and hybrid quantum-classical methods. These strategies concern how a simulation is performed and can be implemented on different hardware platforms. For a broader discussion of quantum simulation architectures, their capabilities, and the challenges of characterizing and validating their performance, we refer the reader to e.g.~\cite{Altman:2019vbv}.

In essence, analog quantum simulators are controllable quantum systems whose natural Hamiltonian can be tuned to reproduce the dynamics of another quantum model. Their analog character comes from the fact that one engineers the dynamics of the quantum platform to emulate those of the target physical system directly, rather than decomposing the evolution into a universal sequence of elementary operations, which is characteristic of digitized approaches. Examples of such platforms include ultracold atoms in optical lattices, trapped ions, superconducting circuits, and Rydberg-atom arrays, among many others~\cite{BlochDalibardNascimbene2012,GeorgescuAshhabNori2014}. The strength of analog simulators is that they use the native dynamics of the device itself. When the available degrees of freedom and interactions closely match those of the target model, the desired evolution is generated directly by the physical Hamiltonian, and this makes analog platforms especially powerful for exploring large many-body systems on near-term devices. Indeed, current experimental set-ups can already realize systems containing tens to hundreds of effective degrees of freedom, see e.g.~\cite{Bernien2017,Zhang2017,Ebadi2021}.

\begin{figure}[h!]
    \centering
    \includegraphics[width=.8\textwidth]{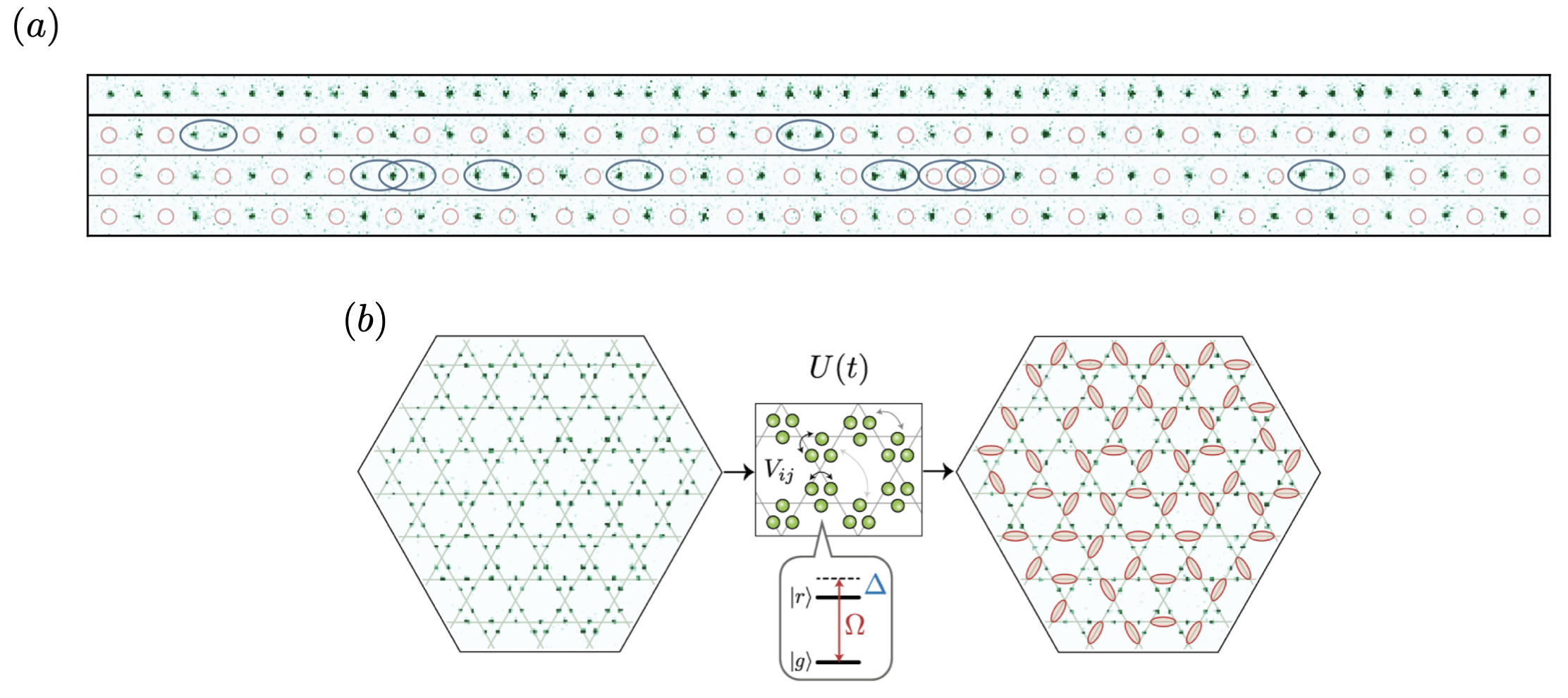}
    \caption{
Rydberg atom arrays for analog simulation of constrained quantum many-body systems.
$(a)$ Fluorescence images of a one-dimensional array of 51 atoms: the top row shows the initial ground-state array, while the lower rows show representative configurations following laser-driven evolution. Green spots indicate ground-state atoms, and red circles mark Rydberg excitations detected through atom loss.
$(b)$ An array of 219 atoms positioned on the links of a kagome lattice, before and after evolution under the Rydberg Hamiltonian. Red bonds represent Rydberg excitations interpreted as dimers, whose allowed configurations are constrained by the blockade.
Figures adapted from~\cite{Bernien:2017ubn,Semeghini:2021wls}, respectively.}
\label{fig:rydberg_gauge}
\end{figure}

As an illustrative example, we consider here an analog platform constructed from neutral atoms. In Rydberg-atom arrays, atoms are trapped at fixed positions, see Fig.~\ref{fig:rydberg_gauge} $(a)$, having two internal atomic states, which can be used to define an effective spin-$1/2$ degree of freedom. The interactions between these spin states are engineered by external lasers. Importantly, the laser driving can be constructed such that the atomic ground state directly couples to the highly excited Rydberg state. Such excited states are interesting due to Rydberg blockade mechanism, which prevents two atoms within a certain radius from being simultaneously in the excited Rydberg state. This is illustrated in Fig.~\ref{fig:rydberg_gauge}$(a)$, where an array of 51 atoms is initially prepared with all atoms in their electronic ground state (green spots, top row). Laser driving couples the ground and Rydberg states, while strong interactions suppress simultaneous Rydberg excitations of nearby atoms through the blockade mechanism. After measurement, atoms in Rydberg states appear as empty sites in the fluorescence image. The resulting excitation patterns reveal spatial correlations induced by the blockade. Identifying the ground and Rydberg states with the two states of an effective spin-$1/2$, the dynamics is described by the Hamiltonian~\cite{Bernien:2017ubn}
\begin{equation}\label{eq:Ryd}
    H_{\rm Ryd}
    =
    \sum_i
    \left(
    \frac{\Omega_i}{2}\sigma_i^x
    -
    \Delta_i n_i
    \right)
    +
    \sum_{i<j} V_{ij} n_i n_j \, ,
\end{equation}
where $n_i$ measures whether atom $i$ is in the excited Rydberg state, $\Omega_i$ is the Rabi frequency, $\Delta_i$ is the laser detuning parameter, and $V_{ij}$ is the interaction between excited atoms. Indeed, Eq.~\eqref{eq:Ryd} gives a clear indication how the native structure of the atomic chain can be mapped to theories with local symmetries: the blockade interaction, $\sum_{i<j} V_{ij} n_i n_j$,
forbids certain nearby excitation patterns, and therefore excludes such states from the Hilbert space which can be explored. With a suitable encoding, these constrained configurations can realize the physical subspace of a lattice gauge theory, providing a way to emulate gauge theories using an atomic array, see e.g.~\cite{SuraceMazzaGiudiciLeroseGambassiDalmonte2019,NotarnicolaColluraMontangero2020,GonzalezCuadra2025} for further discussion. A related realization is illustrated in Fig.~\ref{fig:rydberg_gauge} $(b)$, where atoms placed on the links of a kagome lattice encode dimers through their Rydberg excitations. The Rydberg blockade suppresses configurations in which multiple dimers touch the same vertex. This can lead to a realization of a local Gauss-law constraint~\cite{Semeghini:2021wls}.

Similar constructions can be implemented in other quantum analog platforms, although the microscopic realization of the constraints and effective interactions is necessarily device dependent. More broadly, we note that analog simulation need not always aim at reproducing a target Hamiltonian in all microscopic details. It can also be used to emulate dynamical mechanisms that are shared by different physical systems. We give an example of this in the coming section.

One of the main limitations of analog simulators is that they are typically not universal. Each target model requires a dedicated, and often non-trivial, mapping between the theory's degrees of freedom and those available in the experimental platform. Moreover, as in classical analog computation, the output can be sensitive to errors present in the system and quantum decoherence effects, all of which can drive the realized dynamics away from the target theory. Correcting these effects is generally difficult precisely because the computation is encoded directly in the physical dynamics of the device. Thus, although current analog platforms provide powerful testbeds for quantum simulation, it is natural to seek platforms that are programmable in a universal sense, and where noise can be controlled systematically. This motivates the use of digital quantum computers, which we follow to discuss.

Digital quantum computers can be typically described within the quantum circuit model. In this framework, one assumes to have a quantum device where the elementary degrees of freedom are qubits, i.e. theoretical spin-$1/2$ degrees of freedom, and computations are
performed by applying a sequence of elementary unitary operations, or quantum
gates, on these; see e.g.~\cite{DiVincenzo:2000tra} for general criteria a digital computer should satisfy. These quantum gates correspond, in standard quantum mechanics, to unitary (linear) operators, and they can act on single qubits, on multiple qubits, or on a register of qubits controlled by the state of other qubits. A quantum
algorithm is therefore specified by a circuit, i.e. an ordered sequence of quantum gates, applied on an array of qubits initialized in a controllable computational state, followed by measurements of these. This construction is somewhat similar to the TN formulation, except that here the information being stored in the circuit is its own wavefunction, i.e. information is now quantum, and therefore, to extract any observable one must run the circuit several times to compute expectation values of operators.

The power of this formulation comes from its inherent notion of universality: a suitable set of elementary gates can approximate any unitary transformation on the qubit register to arbitrary accuracy. In a real quantum device, only a finite set of native one- and two-qubit gates is available, and all other operations must be constructed from them. The precise set of gates depends on the hardware being considered to realize the quantum computer. The reason why such a small set is sufficient for universal computation can be easily understood. Single-qubit gates allow one to perform arbitrary rotations of each local two-level system, i.e.
to generate $\mathrm{SU}(2)$ on each qubit. Entangling two-qubit gates
allow information stored in different qubits to become genuinely nonlocal.
Together, these operations give access not only to local rotations, but also to
multi-qubit correlations. A standard finite universal gate set is $\{H,T,\mathrm{CNOT}\}$, where
\begin{equation}
H=\frac{1}{\sqrt{2}}
\begin{pmatrix}
1&1\\
1&-1
\end{pmatrix},
\qquad
T=
\begin{pmatrix}
1&0\\
0&e^{i\pi/4}
\end{pmatrix},
\qquad
\mathrm{CNOT}=
\begin{pmatrix}
1&0&0&0\\
0&1&0&0\\
0&0&0&1\\
0&0&1&0
\end{pmatrix}\, ,
\end{equation}
acting on local one and two qubit Hilbert spaces in the $z$-basis. Here, the CNOT gate is written in the computational basis $\{|00\rangle,|01\rangle,|10\rangle,|11\rangle\}$, with the first qubit acting as the control, i.e. it negates the second qubit if the first is in the $|1\rangle$ state and can generate entanglement. The Hadamard gate changes between the $\sigma^z$ and $\sigma^x$ bases, while the $T$ gate introduces a relative phase between the computational basis states. Repeated application of $H$ and $T$ generates a dense set of single-qubit rotations, up to an overall phase, while CNOT extends this control to entangled many-qubit states.\footnote{The non-Clifford character of $T$ is essential here. Circuits composed of Clifford gates, generated by $H$, $T^2$, and CNOT, acting on stabilizer input states and followed by Pauli measurements, admit efficient classical simulation through the Gottesman--Knill theorem~\cite{Gottesman:1998hu}. Adding $T$ makes this gate set universal.} In practice, universality is not sufficient to have a digital quantum device, as the number of gates needed to approximate a desired operation also determines whether it can be implemented efficiently. This aspect is addressed by the Solovay--Kitaev theorem~\cite{Kitaev1997}, which guarantees that for a fixed number of qubits and a finite universal gate set, a target gate can be approximated to accuracy $\varepsilon$ using a sequence whose length grows only polylogarithmically with $1/\varepsilon$~\cite{Dawson:2005blj}. As a result, once a target Hamiltonian has been encoded in terms of qubits, its real-time evolution can therefore be approximated by a quantum circuit. However, one should bear in mind that whether this can be done efficiently depends on the structure of the Hamiltonian and the particular problem at study.

Besides the flexibility of digital quantum devices in implementing different target models using the basic elements, a second advantage is its compatibility with quantum error correction and fault-tolerant gate constructions, which provides a route towards reliable computation. The gate-based structure allows, in principle, for potential errors generated during computations to be corrected more systematically than is generally possible in analog devices. The basic idea in digital quantum error correction lies in encoding logical qubits into a larger space of many physical
qubits, in analogy to what is done with classical devices. Correctable errors can be detected through syndrome measurements without
directly collapsing the encoded quantum
state. This notion is formalized under the Knill--Laflamme condition~\cite{Knill:1997mom}, which gives  the precise criterion under which a set of errors is correctable, providing the formal basis for fault-tolerant digital quantum computation, see e.g.~\cite{Gottesman:2009zvw,Girvin:2021txg,Spencer:2026mqt} for a more detailed discussion on quantum error correction.

To better elucidate the formulation of digital quantum computers, suppose that a target Hamiltonian has been mapped to an operator (acting on a qubit array) of the form
\begin{equation}\label{eq:eeeaa}
H = \sum_{\alpha} c_{\alpha} P_{\alpha},
\end{equation}
where the $P_{\alpha}$ are tensor product strings of Pauli operators, see Eq.~\eqref{eqPaulis}. Focusing on real-time evolution, the key challenge is to implement the evolution operator $\exp(-iHt)$ in terms of a sequence of quantum gates. In general, unless $H$ is trivial, the time evolution operator is not directly a native quantum gate available in a quantum platform, i.e. a gate operation which the device can directly execute. A standard approach is therefore to use a product formula, usually under the Trotter-Suzuki form~\cite{Trotter:1959ytf,Suzuki:1976be}. Dividing the
time evolution into $N_T$ steps of size $\delta t = t/N_T$, one can write
\begin{equation}\label{eq:eee}
e^{-iHt}
=
\left(e^{-iH\delta t}\right)^{N_T}
\simeq
\left(
\prod_{\alpha} e^{-i c_{\alpha} P_{\alpha} \delta t}
\right)^{N_T}
+ \mathcal{O}(t\,\delta t)\,.
\end{equation}
Here the remainder denotes the accumulated first-order product-formula error at fixed $t$, with a prefactor depending on Hamiltonian commutators. Higher order formulas can be constructed at the cost of a more complex operator structure. Equation~\eqref{eq:eee} ensures that the time evolution operator can be implemented by applying a sequence of simpler operators $e^{-i c_{\alpha} P_{\alpha} \delta t}$, for short time intervals $\delta t$. Importantly, each one of these operators is written as the exponential of a Pauli string, and thus its exponentiation is trivial in terms of Pauli operators. By this we mean, for example, a two-qubit interaction of
the form $\sigma^z_i \otimes \sigma^z_j$ can be implemented by entangling the two qubits, applying a
single-qubit rotation, and then undoing the entangling operation:
\begin{equation}
e^{-i \delta t \sigma^z_i \otimes \sigma^z_j}
=
\mathrm{CNOT}_{ij}
\,
R_z^{(j)}(2\delta t)
\,
\mathrm{CNOT}_{ij} \, ,
\end{equation}
where $\mathrm{CNOT}_{ij}$ denotes a CNOT gate with qubit $i$ as
the control and qubit $j$ as the target, and
$R_z^{(j)}(\delta t)=\exp(-i\delta t \sigma^z_j/2)$ is a single-qubit rotation acting on qubit $j$. The role of the first CNOT is to map the parity of the two-qubit state onto the target qubit, so that the subsequent $R_z$ rotation applies a
phase conditioned on the eigenvalue of $\sigma^z_i\otimes \sigma^z_j$. The second CNOT then uncomputes this parity map, restoring the computational basis while leaving behind the desired two-qubit phase. More general Pauli strings can be treated in the same way.

This construction provides a systematic strategy for implementing Hamiltonian evolution once the Hamiltonian has been decomposed into Pauli strings. However, a straightforward implementation can be costly: each exponential is applied separately at every Trotter step, so the gate count grows with both the number of terms and the number of steps required to control the approximation error. This can produce long (deep) circuits even for relatively simple models, making accurate execution challenging on real noisy devices. Practical digital quantum simulation therefore benefits from optimized product formulas and compilation strategies that exploit the Hamiltonian's structure to reduce circuit costs; see e.g.~\cite{Childs:2019hts,Campbell:2019fez,Low:2016znh,Low:2016sck,Gilyen:2018khw,Martyn:2021eaf} for further discussion on this.

Many digital quantum processors still operate in the noisy intermediate-scale quantum (NISQ) regime~\cite{Preskill2018}, although early demonstrations of quantum error correction are beginning to move beyond it~\cite{GoogleQuantumAIandCollaborators:2024efv,Bluvstein:2023zmt}. Representative superconducting and trapped-ion processors contain $\mathcal{O}(10^2)$ physical qubits, while neutral-atom arrays can contain several thousand atoms~\cite{GoogleQuantumAIandCollaborators:2024efv,Helios2025,Manetsch2025}. These physical-system sizes should not be confused with the number of available error-corrected logical qubits. As a rough estimate, assuming independent stochastic errors, a circuit containing $G$ gates with error probability $p_{\rm eff}$ per gate has a no-error probability
\(
P_0=(1-p_{\rm eff})^G\simeq e^{-G p_{\rm eff}}.
\)
For illustrative two-qubit error probabilities $p_{\rm eff}\sim10^{-3}$--$10^{-2}$, the probability of at least one error becomes appreciable after $G\sim10^2$--$10^3$ entangling gates. Although this is not a strict limit on useful computation, it illustrates how accumulated noise restricts accessible circuit depths and simulation times. For a broader discussion of experimental progress across quantum computing platforms, see~\cite{LeRegent2025}.

Finally, the limitations of NISQ era quantum devices have strongly motivated hybrid quantum-classical approaches, which delegate only part of the
computation to the quantum device, while using classical resources to optimize and control the calculation. A paradigmatic example is the variational quantum
eigensolver (VQE) algorithm~\cite{Peruzzo:2013bzg}. In this algorithm a quantum processor prepares a parametrized trial quantum state
$|\psi(\boldsymbol{\theta})\rangle$, using an ansatz typically chosen to keep the circuit depth manageable. The energy of the state can be obtained from $
E(\boldsymbol{\theta})
=
\langle \psi(\boldsymbol{\theta})|H|\psi(\boldsymbol{\theta})\rangle$, 
and is estimated through repeated measurements of the terms in the Hamiltonian. A classical optimizer is then used to update the parameters $\boldsymbol{\theta}$ (which characterize the sequence of quantum gates) so as to minimize the state energy, and the procedure is iterated until convergence. Such algorithms have been widely explored in quantum chemistry and condensed matter applications~\cite{Tilly:2021jem}.

\subsection{Mapping gauge theories to spin models: Schwinger model example}

To conclude this section, we provide an explicit example of how to map a lattice
gauge theory, the most basic formulation underlying nearly all relevant problems in NP, to a system whose basic degrees of freedom are spin-$1/2$ variables.
We shall consider quantum electrodynamics in one spatial dimension, more commonly
known as the Schwinger model~\cite{Schwinger:1962tp,Schwinger:1962tn,Lowenstein:1971fc,Coleman:1975pw,Coleman:1976uz}, which although  considerably simpler than
QCD, provides a useful starting point to discuss the simulation of gauge
theories. This model has been widely studied in recent years within the QIS context, see e.g. for some recent examples~\cite{Klco:2021lap,Halimeh:2023lid,Davoudi:2021ney,Banerjee:2012pg,Hauke:2013jga,Rico:2013qya,
Pichler:2015yqa,Martinez:2016yna,Muschik:2016tws,
Klco:2018kyo,Kokail:2018eiw,Surace:2019dtp,
Mil:2019pbt,Davoudi:2019bhy,Kharzeev:2020kgc,
Shaw:2020udc,Yang:2020yer,Rigobello:2021fxw,
Chakraborty:2020uhf,Nguyen:2021hyk,Zhou:2021kdl,
Zache:2021ggw,Halimeh:2022pkw,Mueller:2022xbg,
Florio:2023dke,Belyansky:2023rgh,Farrell:2023fgd,
Farrell:2024fit,Barata:2023jgd,Florio:2024aix,
Zhang:2023hzr,Barata:2025hgx}. More generally, there has been a very active effort in the first-principles formulation of lattice gauge theories in the Hamiltonian picture, i.e. the natural picture for quantum devices, where the biggest challenge lies in incorporating the notion of local gauge symmetry in a computationally and resource efficient manner, see e.g.~\cite{Rogerson:2026jjt,
Raychowdhury:2019iki,Raychowdhury:2018osk,
Kadam:2022ipf,Kadam:2024ifg,
Davoudi:2020yln,Zohar:2018cwb,Bender:2020ztu,
Pardo:2022hrp,Haase:2020kaj,Ciavarella:2021nmj} for recent developments.

The Schwinger model contains several characteristic
features of higher-dimensional gauge theories, including confinement, charge screening, a
non-trivial vacuum structure, and a chiral anomaly, see~\cite{Coleman:1975pw,Schwinger:1962tn}, which have a  different origin compared to QCD. In the temporal gauge, i.e. $A^0=0$, the
continuum Hamiltonian of the massive Schwinger model can be written as
\begin{equation}
H
=
\int dx
\left[
\bar{\psi}(x)
\left(
-i\gamma^1 D_1 + m
\right)
\psi(x)
+
\frac{1}{2}E^2(x)
\right],
\label{eq:schwinger_continuum}
\end{equation}
where $\psi$ is a charged fermion field,
$D_1=\partial_1+igA_1$ is the gauge-covariant derivative, and $E$ is the
electric field. The first term describes the motion and mass of the charged
matter field, while the second term is the electric field energy. There is no
magnetic field term, since in one spatial dimension there is no magnetic flux.
Moreover, the electric field is constrained by Gauss' law,
\begin{equation}
    \nabla E = g \psi^\dagger \psi .
\end{equation}
Thus, in one dimension the gauge field does not describe an independent
propagating degree of freedom. Instead, it acts as a constrained field
determined by the matter charge distribution.

We now wish to find a lattice representation of this theory. To this end, we discretize
space into a one-dimensional lattice with spacing $a$. A naive discretization
of the fermion derivative, however, does not produce a single relativistic
fermion in the continuum limit. Instead, the lattice dispersion relation
develops additional low-energy modes near the edges of the Brillouin zone,
which appear as extra fermion species. This corresponds to the so called fermion-doubling problem.
To reduce this redundancy in one spatial dimension, we use the
Kogut--Susskind staggered-fermion prescription~\cite{Kogut:1974ag,Banks:1975gq,Susskind:1976jm}, which is one among several prescriptions~\cite{Wilson:1975id,Drell:1976mj,Kaplan:1992bt,
Shamir:1993zy,Ginsparg:1981bj,Neuberger:1997fp} to remove the edge modes.
In this formulation the components of the Dirac spinor are distributed over
neighboring lattice sites, so that a single one-component fermion field
$\chi_n$ lives on each site. The lattice Hamiltonian is then~\cite{Hamer:1997dx}
\begin{equation}
H_{\rm lat}
=
-\frac{i}{2a}
\sum_n
\left(
\chi_n^\dagger U_n \chi_{n+1}
-
\chi_{n+1}^\dagger U_n^\dagger \chi_n
\right)
+
m
\sum_n
(-1)^n
\chi_n^\dagger \chi_n
+
\frac{a g^2}{2}
\sum_n
L_n^2 .
\label{eq:schwinger_lattice}
\end{equation}
Here $U_n$ is the
gauge link connecting sites $n$ and $n+1$, and $L_n=E_n/g$ is the corresponding
electric field operator. The link operator $U_n$ shifts the electric flux and
satisfies
\begin{equation}
[L_n,U_m]=\delta_{nm}U_n .
\end{equation}
The hopping term in Eq.~\eqref{eq:schwinger_lattice} moves a charged particle
between neighboring sites, but it does so together with the gauge link. In this
way, when a charge moves, the electric flux changes consistently with charge conservation. The physical Hilbert space is selected by Gauss' law. For open boundary
conditions, the local generator of gauge transformations can be written as
\begin{equation}
G_n
=
L_n-L_{n-1}
-
\chi_n^\dagger \chi_n
+
\frac{1-(-1)^n}{2}\, ,
\label{eq:gauss_law}
\end{equation}
where the last term in Eq.~\eqref{eq:gauss_law} is the staggered background charge,
which appears because the Dirac sea is represented by alternating occupied and
empty sites. Physical states must satisfy
\begin{equation}
G_n|\Psi\rangle=0
\qquad
\text{for all } n .
\end{equation}
Thus, to simulate the theory one must ensure that the computational state remains in this gauge-invariant subspace.

In one spatial dimension, Gauss' law can be solved recursively. Numbering the sites from $n=0$ and fixing the dimensionless electric flux at the left boundary to be $L_{-1}=\ell$, the electric flux on link $n$ is determined by the total charge to its left,
\begin{equation}
L_n
=
\ell
+
\sum_{j=0}^{n}
\left[
\chi_j^\dagger \chi_j
-
\frac{1-(-1)^j}{2}
\right].
\label{eq:solved_gauss_fermions}
\end{equation}
Thus, once the matter configuration is known, the electric field is fixed. This
allows one to rewrite the Hamiltonian entirely in terms of the fermionic matter
fields, at the price of a long-range
Coulomb interaction among the charges. After eliminating the spatial gauge links by a gauge choice, one can map the resulting fermionic chain to spins. This step is non-trivial because fermionic operators
anticommute, while spin operators on different sites commute. A standard way to
recover the correct fermionic algebra is to apply a Jordan Wigner transformation~\cite{Jordan:1928wi}
\begin{equation}
\chi_n
=
\left(
\prod_{j<n} \sigma_j^z
\right)
(-i)^n\sigma_n^-,
\qquad
\chi_n^\dagger
=
\left(
\prod_{j<n} \sigma_j^z
\right)
i^n\sigma_n^+ \, ,
\label{eq:jordan_wigner}
\end{equation}
where the string of $\sigma^z$ operators ensures the correct fermionic
anticommutation relations, as can be explicitly checked. The site-dependent phases make the hopping term real in the spin representation. Using
$
\chi_n^\dagger \chi_n
=
\frac{1+\sigma_n^z}{2}$, 
one obtains the spin Hamiltonian
\begin{equation}
H_{\rm spin}
=
\frac{1}{2a}
\sum_n
\left(
\sigma_n^+ \sigma_{n+1}^-
+
\sigma_n^- \sigma_{n+1}^+
\right)
+
\frac{m}{2}
\sum_n
(-1)^n \sigma_n^z
+
\frac{a g^2}{2}
\sum_n
\left[
\ell
+
\frac{1}{2}
\sum_{j=0}^{n}
\left(
\sigma_j^z+(-1)^j
\right)
\right]^2 ,
\label{eq:schwinger_spin}
\end{equation}
up to additive constants. In this form,
gauge invariance is already built into the Hamiltonian through the electric field
term. The resulting model is therefore directly suitable for TN
or digital quantum computer based implementations.

Although the previous strategy works well in one spatial dimension, in higher dimensions solving Gauss' law generally produces complicated non-local
interactions and one can not fully integrate out the gauge field. As a result, alternative descriptions keeping track of the gauge degrees of freedom should also be considered. Since the electric field of a compact $U(1)$ gauge theory has an
infinite-dimensional Hilbert space, one must then introduce a finite-dimensional
truncation. Here we briefly discuss one widely used representation, the so called quantum link model formulation~\cite{Horn:1981kk,Orland:1989st,Chandrasekharan:1996ih,Wiese:2021djl}. In this approach the link
operators are represented by spin-$S$ operators,
\begin{equation}
L_n \to S_n^z,
\qquad
U_n \to \frac{S_n^+}{\sqrt{S(S+1)}},
\qquad
U_n^\dagger \to \frac{S_n^-}{\sqrt{S(S+1)}} .
\label{eq:qlm_map}
\end{equation}
The Hamiltonian in Eq.~\eqref{eq:schwinger_lattice} becomes
\begin{equation}
H_{\rm QLM}
=
-\frac{i}{2a}
\sum_n
\left(
\chi_n^\dagger S_n^+ \chi_{n+1}
-
\chi_{n+1}^\dagger S_n^- \chi_n
\right)
+
m
\sum_n
(-1)^n
\chi_n^\dagger \chi_n
+
\frac{a g^2}{2}
\sum_n
(S_n^z)^2 ,
\label{eq:schwinger_qlm}
\end{equation}
up to the normalization of the link operator. Gauge invariance is now imposed by
the local constraint
\begin{equation}
G_n
=
S_n^z-S_{n-1}^z
-
\chi_n^\dagger \chi_n
+
\frac{1-(-1)^n}{2},
\qquad
G_n|\Psi\rangle=0 .
\label{eq:qlm_gauss}
\end{equation}
In an implementation with gauge-violating perturbations, one may energetically suppress
gauge-violating states by adding a penalty term,
\begin{equation}
H'
=
H_{\rm QLM}
+
\lambda
\sum_n G_n^2,
\qquad
\lambda \gg \max\{a^{-1},|m|,ag^2\} .
\label{eq:gauge_penalty}
\end{equation}
Here $\lambda$ is chosen large compared with the relevant Hamiltonian energy scales, as finite penalties suppress rather than exactly remove gauge violations.
The quantum link model therefore keeps the local gauge-field structure explicit,
while replacing each link by a finite-dimensional Hilbert space. Such a formulation is better suited for real quantum devices, where, for example, connectivity between qubits is typically limited, while for TN methods this formulation is also useful, as it avoids an all-to-all potential term in the Schwinger model.

\section{Near-term applications in nuclear physics}\label{sec:applciation_NP}
In the previous two sections, we first reviewed a set of open problems in high energy nuclear physics and explained why the limitations of the existing theoretical and computational tools hinder progress in answering these questions. We then introduced several tools from quantum information science that can address some of these bottlenecks, including real-time dynamics and the simulation of quantum systems with large Hilbert spaces. In this section, we illustrate how these methods can be directly applied to problems closely connected to high energy nuclear physics. We present several representative examples drawn from the literature and provide additional references for interested readers. 

\subsection{Particle scattering, fragmentation and hadronization}\label{subsec:scattering}

One of the early seminal discussions on applying quantum simulation to study quantum field theory was put forward by Jordan, Lee, and Preskill~\cite{Jordan:2012xnu,Jordan:2011ci,Jordan:2014tma}. These works showed how (real-time) $S$-matrix problems, see Eq.~\eqref{eq:Smatrix}, can be reformulated as a quantum algorithm, providing a \textit{dynamical} formulation of the scattering process in a digital quantum device. The conceptual novelty of these works was therefore not only to propose a digital simulation of a field theory, but to give an end-to-end procedure for computing scattering amplitudes in regimes where perturbation theory need not be applicable, as discussed above.\footnote{ See e.g.~\cite{Briceno:2026hbx,Burbano:2025pef,Li:2023kex,Turco:2023rmx,Turro:2024ksf} for other approaches to computing scattering amplitudes in quantum devices, and realizations and extensions of the Jordan, Lee, Preskill program\cite{Marshall:2015mna,Barata:2020jtq,Liu:2021otn,
Kreshchuk:2023btr,Chai:2023qpq,Farrell:2024fit,Davoudi:2024wyv,
Hardy:2024ric,Zemlevskiy:2024vxt,Abel:2025zxb,
Farrell:2025nkx,Ingoldby:2025bdb}.}

At present, however, realizing this full $S$-matrix program relevant for collider phenomenology remains far beyond near/mid-term capabilities of quantum devices, with several challenges of different origin: the incoming hadronic states are themselves highly non-trivial bound states~\cite{Davoudi:2024wyv,Turco:2023rmx,Bennewitz:2025nhz,Farrell:2024fit}, the real-time evolution must resolve a large hierarchy of physical scales, and the final state contains many particles whose inclusive probabilities must be reconstructed from measurements, see e.g.~\cite{Barata:2026aoq,Barata:2024apg,Cao:2026nyj}. Moreover, a crude resource estimate indicates that this task requires substantial computational resources: in a hadronic collision with a hard scale $Q$, the lattice spacing $a$ must be small enough to resolve momenta of order $Q$, $a\lesssim Q^{-1}$, while the physical volume and evolution time must be large enough to capture dynamics at hadronic scales of order $\Lambda_{\rm QCD}$. Even for $Q\sim 1~{\rm TeV}$ and $\Lambda_{\rm QCD}\sim 0.2$--$1~{\rm GeV}$, this requires $a\lesssim 2\times10^{-4}~{\rm fm}$ and implies a dynamical range $Q/\Lambda_{\rm QCD}\sim 10^3$--$5\times10^3$, corresponding to at least $\mathcal{O}(10^9$--$10^{11})$ spatial lattice sites in three dimensions for a box of size $L\gtrsim\Lambda_{\rm QCD}^{-1}$. This hierarchy becomes still larger for higher hard scales accessible at, e.g.,  the LHC. A direct simulation of the full scattering process would therefore require a very fine lattice over a large volume.

This disparity in scales motivates a more modest near-term strategy: rather than simulating a full collider event from the incoming hadrons to the final detected particles, one can isolate particular (real-time) components of the problem that are difficult for standard methods, see~\cite{Bauer:2021gup} for related discussion. In spirit, this idea fits well into the scale-dependent factorization in QCD, where processes occurring at different scales are treated separately with different methods. In the present context, this separation is valuable as it reduces the portion of the problem that must be treated on a quantum device, thus requiring fewer resources. We further note that this strategy is sensible as for example a hard process in QCD, i.e. dominated by a very energetic scale, can be well described in perturbation theory, while softer physical processes are better described using lattice/non-perturbative methods. Thus, in what follows, we show examples where (sub-)parts of a full $S$-matrix scattering picture are treated using QIS methods.

\begin{figure}[h!]
\centering
\includegraphics[width=1\textwidth]{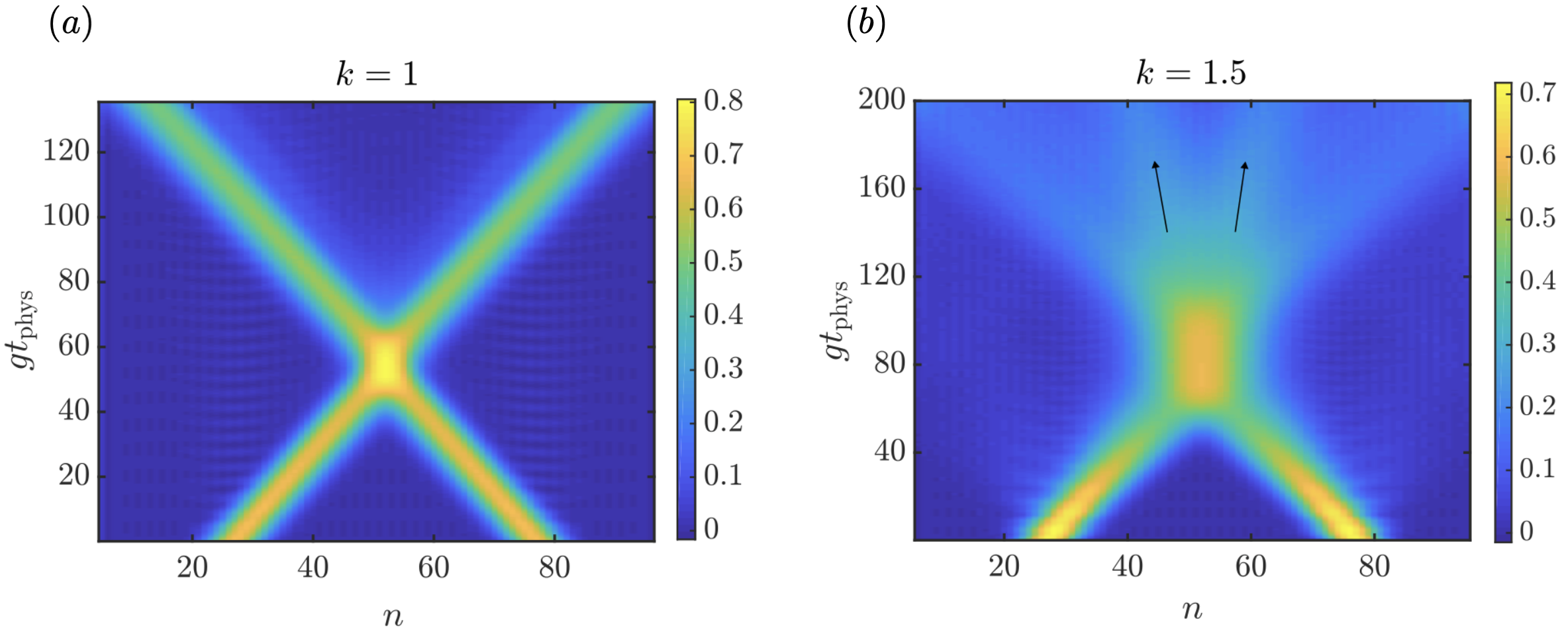}
\caption{Real-time meson-meson scattering in the lattice Schwinger model from tensor network simulations. The color maps show the excess two-site entanglement entropy, with the vacuum contribution subtracted, as a function of lattice position $(n)$ and physical time $g t_{\rm phys}$. $(a)$ For incoming vector mesons with momentum $k=1$, the collision remains below the inelastic threshold and the dominant signal corresponds to elastic scattering, with two outgoing mesons propagating away from the collision region. $(b)$ For $k=1.5$, the incoming mesons have sufficient energy to open an inelastic channel. In addition to the outgoing elastic components, a slower signal appears near the center of the system, corresponding to the production of two heavier scalar bound states, the arrows indicate their trajectories. Figure adapted from~\cite{Papaefstathiou:2024zsu}.}
\label{fig:scattering_Bahuls}
\end{figure}

As a first illustration, in Fig.~\ref{fig:scattering_Bahuls} we show a TN based computation where the scattering of two vector mesons in the lattice Schwinger model is studied~\cite{Papaefstathiou:2024zsu}. In the parameter regime considered, the spectrum of the theory contains a vector meson and a heavier scalar meson, below the multi-particle continuum. After preparing two wave-packets of the vector meson (yellow bands at $t=0$ in both panels) with opposite momenta, these are let to dynamically evolve and scatter, while the outgoing states are measured for a finite time interval. The two panels differ by the initial momenta $k$ of the wave-packets, in panel $(a)$ $k$ is set below the particle threshold for producing the scalar meson, while panel $(b)$ takes $k$ above the production threshold. Below the kinematic threshold, the dominant process is elastic scattering: the two vector mesons collide and then propagate away from the interaction region, with slight broadening of the wave-packets. Above threshold, the inelastic channel opens for the production of the heavier scalar bound states. This is reflected in the shift of the entropy distribution peaks, which are now centered in the middle of the lattice, as one expects for the production of slowly moving heavier states. Their evolution can be further analyzed using other observables~\cite{Papaefstathiou:2024zsu, Artiaco:2025qqq}, and their velocities are indicated by the arrows. As a side note, this computation provides a nice example on how quantum simulation can go beyond the strict $S$-matrix picture, by allowing to track the full evolution for any time, and not just capturing the properties of asymptotic states. For further examples of particle scattering using real-time methods we refer an interested reader to several recent studies~\cite{Surace:2020ycc,Rigobello:2021fxw,
Belyansky:2023rgh,Su:2024uuc,
Davoudi:2025rdv,Schuhmacher:2025ehh,Chai:2025qhf,
Joshi:2025rha,Xiang:2025qhq,Barata:2025rjb,Barata:2025hgx}.

\begin{figure}[h!]
\centering
\includegraphics[width=1\textwidth]{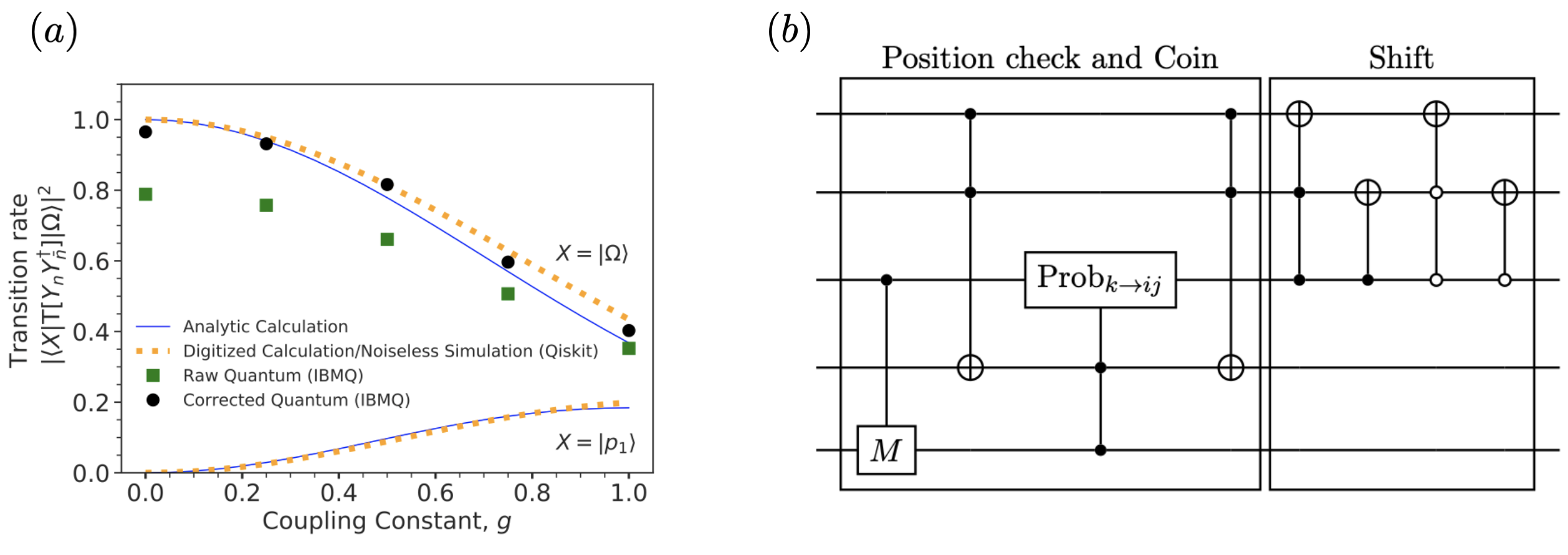}
\caption{Quantum computing approaches to high energy scattering and parton shower dynamics. $(a)$ Transition rates for a simple scalar effective field theory observable as a function of the coupling $g$, comparing the analytic result with a digitized/noiseless Qiskit simulation and raw/error-mitigated IBMQ measurements. $(b)$ Quantum-walk circuit for a parton shower step, in which the position check and coin operations encode the shower branching probabilities, while the conditional shift updates the shower history. Figures adapted from~\cite{Bauer:2021gup,Bepari:2021kwv}, respectively.}
\label{fig:PShowers}
\end{figure}

QIS tools can be more directly combined with the high energy QCD elements discussed in Section~\ref{sec:Nuclear}, see Fig.~\ref{fig:pp_fig} $(b)$. In Fig.~\ref{fig:PShowers}, we show two illustrations of this. In panel $(a)$, a toy demonstration for the computation of a \textit{soft} transition rate in a free scalar theory is shown. This object would enter the (perturbative) computation of a so called soft factor, which in QCD factorization formulas describes the production of soft partonic radiation. The presented results show how an analytic calculation in the free scalar theory, which is the simplest possible case, is not reproduced by a \textit{raw} run on an IBM digital device, requiring dedicated quantum error mitigation strategies. Although this simple example admits an exact closed form solution, the extension of such computations to the QCD context illustrates well the need to combine insights into the theory, i.e. separation of scales and factorization, to reduce the complexity of the quantum computation to be performed, such that it becomes amenable to available platforms, and the importance of controlling the noise in such simulations.

In panel $(b)$, we show a different example of this synergy between QCD phenomenology and quantum simulation techniques. As discussed above, when high energy partons are produced they tend to fragment to a large number of final state particles (see Fig.~\ref{fig:pp_fig}). At a given theoretical accuracy, this cascading process can be described using importance sampling techniques, usually referred to as parton showers, see e.g.~\cite{Buckley:2011ms,Hoche:2014rga,
Dasgupta:2020fwr}. This probabilistic treatment is determined by a (non-branching) probability distribution
\begin{equation}\label{eq:ttt}
\Delta_k(t_{\rm max},t)
=
\exp
\left[
-
\sum_{i,j}
\int_t^{t_{\rm max}}
\frac{dt'}{t'}
\int dz\,
\frac{\alpha_s(t')}{2\pi}
P_{k\to ij}(z)
\right]\, ,
\end{equation}
where $t$ denotes the evolution scale, and $P_{k\to ij}(z)$ is the basic probability weight for a parton $k$ to produce partons $i$ and $j$, with an energy sharing fraction $z$, which is computable in perturbative QCD~\cite{Gribov:1972ri,Gribov:1972rt,Dokshitzer:1977sg,Altarelli:1977zs}. Panel $(b)$ of Fig.~\ref{fig:PShowers} recreates this process in terms of a quantum-walk quantum circuit architecture. At each step, a quantum coin flip decides whether or not a new particle is produced following Eq.~\eqref{eq:ttt}, which leads to an update to quantum registers tracking history of the shower. The circuit coherently represents a superposition of shower histories, while measurement samples a given history. Quantum-walk approaches can offer computational advantages in suitable formulations, while their practical benefit depends on encoding and measurement costs~\cite{Szegedy:2004ony, Bepari:2021kwv}. More interestingly, these quantum strategies can be used to track interference processes, which are hard to capture using classical methods. One example of this, introduced in~\cite{Bauer:2019qxa},  is the case where oscillations between the propagation and flavor bases exist (similar to neutrino oscillations), which can be well accommodated in a quantum-walk, by rotating between bases at each step, while they can not be fully captured using standard importance sampling methods.

Finally, an essential part of any simulation of a high energy scattering event is description of the parton to hadron transition, which for example one would need to implement after running a parton shower simulation. As discussed in Section~\ref{subsec:pp}, the short-distance degrees of freedom produced in a hard scattering are quarks and gluons, while the asymptotic particles observed in the detector are color singlet hadrons. The transition between these two descriptions occurs at scales of order $\Lambda_{\rm QCD}$, where perturbation theory breaks down. In phenomenological event generators this stage is modeled rather than derived from first principles QCD, despite being motivated by current understanding of non-perturbative aspects of the theory, see e.g.~\cite{Bali:2000gf,Luscher:2004ib,Aharony:2013ipa}. The Lund model~\cite{Andersson:1983ia}, an example of such a phenomenological hadronization model, represents the color field between separating partons as a flux tube or string. As the partons separate, the energy stored in the string grows approximately linearly with their separation. When this energy becomes large enough, it becomes favorable to produce a new quark-antiquark pair from the vacuum, breaking the original string into shorter color-connected pieces that eventually form hadrons, see~\cite{Webber:1983if,Buckley:2011ms} for alternative hadronization models in event generators. The microscopic intuition behind string breaking is closely associated to the Schwinger effect, which predicts that a sufficiently strong electric field makes the vacuum unstable to the production of charged particle-antiparticle pairs. Semiclassically, this can be understood as tunneling through the energy barrier created by the electric field, with a (Schwinger) rate schematically of the form $\frac{\Gamma}{L}
\sim
\exp\left[
-\frac{\pi m^2}{g E}
\right]$, where $m$ is the particle mass, $g$ the gauge coupling, and $E$ the electric field. Thus, in a confining theory such as the Schwinger model, the pair production rate depends strongly on the ratio $m^2/(gE)$, while string breaking also depends on the available energy and string geometry.

\begin{figure}[h!]
\centering
\includegraphics[width=1\textwidth]{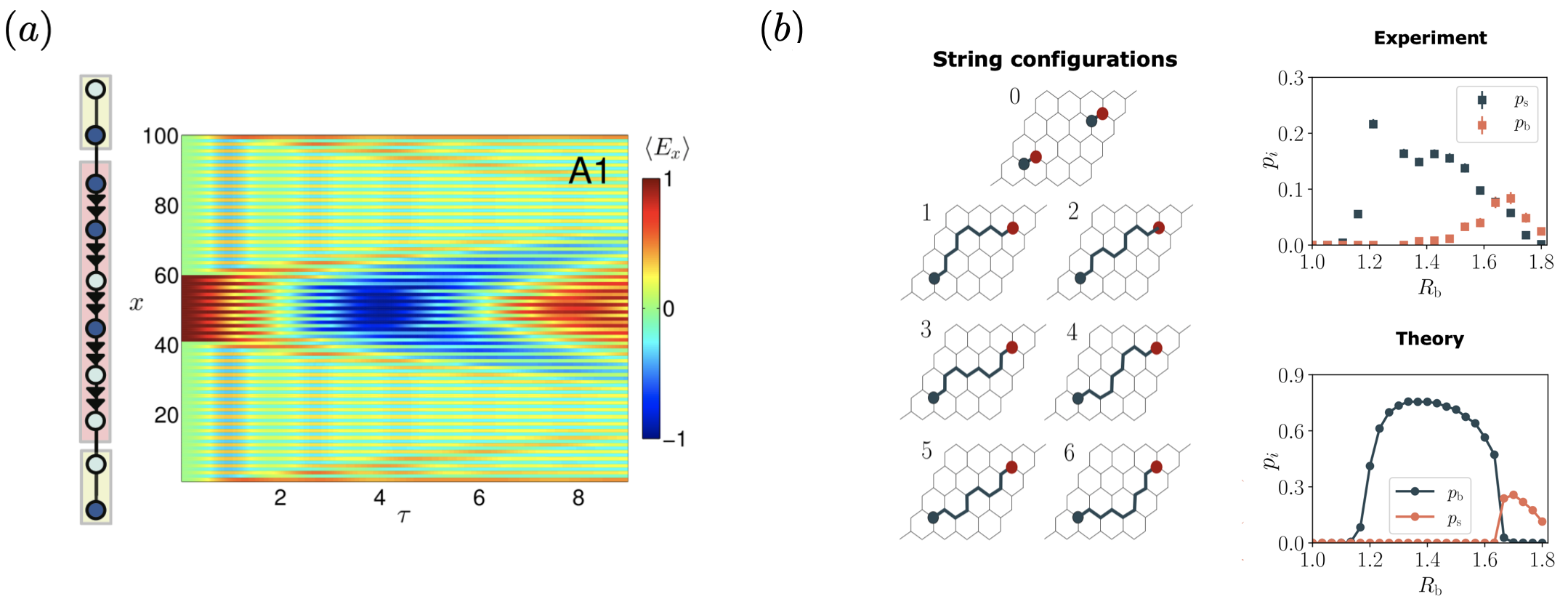}
\caption{Real-time string breaking in lattice gauge theories. $(a)$ Tensor network simulation of a $1+1$-d U(1) lattice gauge theory with dynamical matter, showing how an initially prepared electric string between external charges can break through Schwinger pair production. The real-time evolution of local electric field and matter observables, together with the growth of entanglement, provides a direct diagnostic of the string breaking process. $(b)$ Observation of string breaking in a programmable $2+1$-d Rydberg quantum simulator. Atoms arranged on a kagome geometry realize an effective confining U(1) lattice gauge theory with tunable particle masses and string tension. Prepared string states can be quenched to reveal both intact and broken string configurations. Figures adapted from~\cite{Pichler:2015yqa,Gonzalez-Cuadra:2024xul}, respectively.}
\label{fig:strings}
\end{figure}

QIS methods are thus useful as they allow to study the dynamics of strings in real-time, allowing for the observation of breaking dynamics across parameter space. An early example was presented in~\cite{Pichler:2015yqa}, where TN methods were used to study strings in a $1+1$-d U(1) lattice gauge theory with dynamical matter. Results from this work are shown in Fig.~\ref{fig:strings} $(a)$. Starting from an initial state with an electric string connecting external charges (left column), the real-time evolution tracks the string dynamics by measuring the electric field at each time. The initial string with a maximal electric field polarized upwards, quickly decays, generating a new central string with the opposite polarization. The outer edges show the propagation of bound singlets. This corresponds exactly to the dynamical string breaking mechanism, where the initial $e^+ e^-$ pair evolves to a $e^+ e^- e^+ e^-$ state. This process is then repeated, leading to a novel breaking at $\tau\approx 6.5$. This example thus provides a first principles realization of the dynamics incorporated in e.g. the Lund hadronization model, starting from first-principles simulation of the underlying theory. The dynamics of strings in several gauge theories, have been widely explored by a large number of groups, see e.g.~\cite{
Banerjee:2012pg,Hebenstreit:2013baa,Buyens:2013yza,
Kuhn:2015zqa,Buyens:2015tea,Buyens:2016hhu,Sala:2018dui,
Surace:2019dtp,Chanda:2019fiu,Verdel:2019chj,Verdel:2023mmp,De:2024smi,
Surace:2024bht,Ciavarella:2024lsp,Liu:2024lut,
Mallick:2024slg,Alexandrou:2025vaj,Gupta:2026tcg} for recent examples.

In more recent years, string properties have also been studied beyond one spatial dimensional theories using quantum platforms. We show an example of this from~\cite{Gonzalez-Cuadra:2024xul} in Fig.~\ref{fig:strings} $(b)$. Here a programmable Rydberg atom array was used to realize an effective $2+1$-d U(1) lattice gauge theory with dynamical matter. The Rydberg blockade enforces a local constraint that plays the role of Gauss' law, while the interactions between atoms generate an effective confining potential, as discussed above. The figure shows several possible string configurations between two external charges. The different configurations reflect the extra freedom in higher dimensions which allows transverse string fluctuations and different possible string geometries. The experiment prepares different string configurations and then uses local quenches to observe whether the string remains intact or breaks into screened configurations, as indicated on the right panel. While large-scale simulations with a controlled connection to continuum gauge theories remain demanding, this and related recent studies~\cite{Cochran:2024rwe,
Borla:2025gfs,DiMarcantonio:2025cmf,Xu:2025abo,
Cobos:2025krn,Tian:2025mbv,Cataldi:2025cyo,
Cao:2026qky,Rogerson:2026jjt,Joshi:2026hfe,
Xu:2026ibi,Gandon:2026das,Bacciconi:2026qrs} give valuable information about non-perturbative real-time, out of reach for Euclidean methods.

\subsection{Hadronic structure}\label{subsec:hadronic_structure}

As discussed in Section~\ref{subsec:EIC}, a central obstacle in connecting QCD to high energy scattering data is the first-principles determination of hadronic structure. We showed that, in the factorized descriptions of inclusive processes, the long-distance structure of the incoming hadron is encoded in universal objects such as PDFs. These quantities are non-perturbative, and their operator definitions involve unequal-time correlations, which are hard to access using Euclidean methods. 

Here we consider the simplest example of the quark PDF introduced in Eq.~\eqref{eq:pdf-definition}. Its calculation requires two ingredients. The first is the preparation of the target hadron state, or more generally a gauge-invariant bound state of the theory. In Euclidean lattice QCD this is achieved by inserting interpolating operators and evolving in Euclidean time. In a Hamiltonian or real-time formulation this projection mechanism is no longer available in the same form: real-time evolution produces oscillatory phases rather than exponential damping. As a result, QIS based approaches require different state preparation strategies. For small systems, exact diagonalization provides the entire spectrum and therefore allows one to identify the desired state directly. For larger (one-dimensional) systems, TN methods provide a scalable alternative using DMRG to variationally prepare the vacuum and low-lying excited states in fixed quantum number sectors~\cite{Banuls:2013jaa,Rigobello:2023ype,Banuls:2025wiq}. Beyond two dimensions and/or utilizing quantum platforms one can make use of variational methods as VQE~\cite{Atas:2021ext,Li:2021kcs,Farrell:2024fit}, or approximate the lowest lying states by using some insight into the theory, see~\cite{Papaefstathiou:2024zsu,Davoudi:2024wyv,
Rigobello:2021fxw,Davoudi:2025rdv} for examples. 

\begin{figure}[h!]
\centering
\includegraphics[width=1\textwidth]{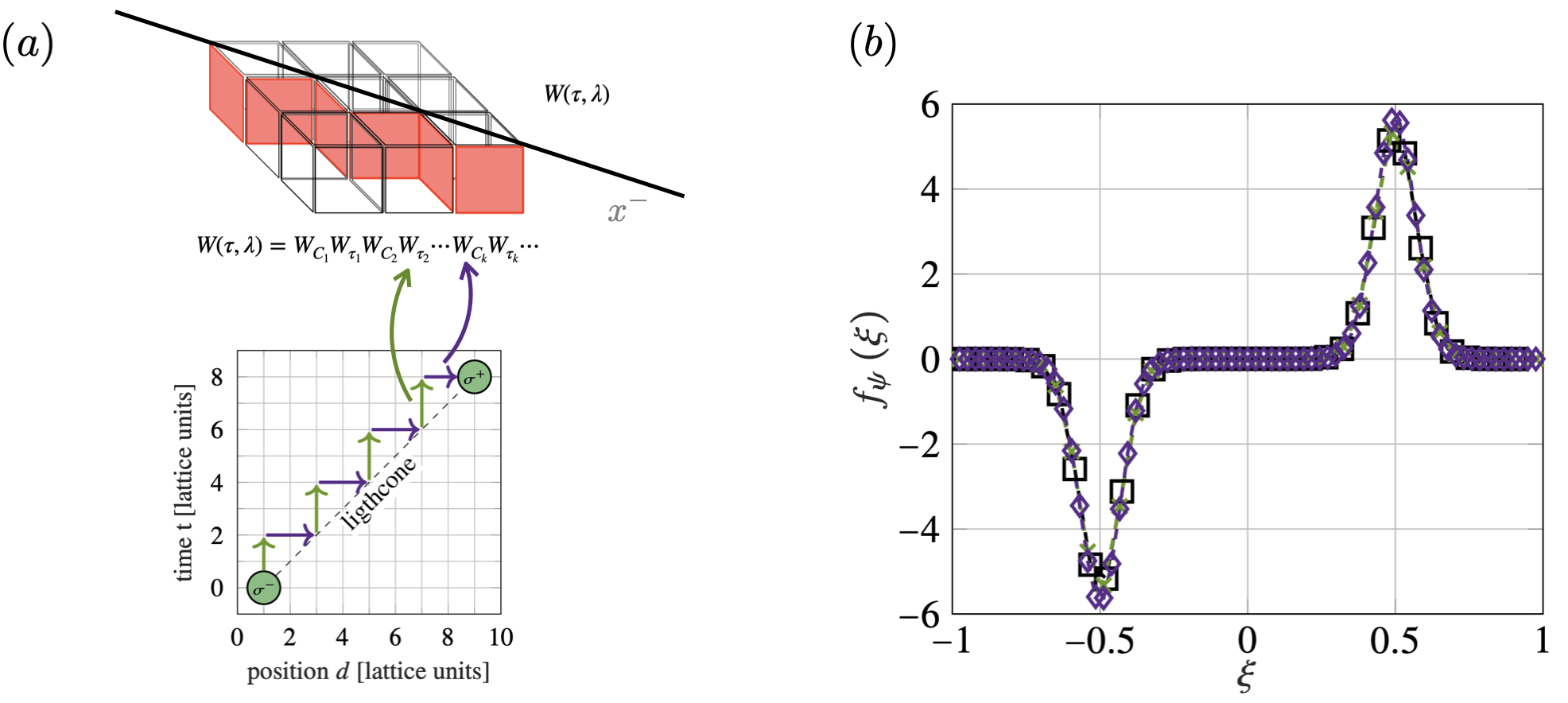}
\caption{Hadronic structure from QIS methods. $(a)$ The light-front Wilson line appearing in PDF definitions can be approximated by a sequence of short real-time evolutions and spatial gauge-string operations, making it suitable for Hamiltonian simulation. $(b)$ Tensor network calculations in the massive Schwinger model then allow the corresponding non-local correlator to be evaluated directly in Minkowski space. Figures adapted from~\cite{Echevarria:2020wct,Banuls:2025wiq}, respectively.}
\label{fig:pdf_quantum}
\end{figure}

We focus here on the second aspect, i.e. the evaluation of the light-like separated operator, which is illustrated in Fig.~\ref{fig:pdf_quantum} $(a)$, following~\cite{Echevarria:2020wct,Banuls:2025wiq}. The PDF correlator contains a Wilson line extending along the lightcone, and therefore combines spatial transport and real-time evolution. To account for both, Ref.~\cite{Echevarria:2020wct} suggests to approximate the Wilson line by a \textit{ladder} made of alternating short real-time and spatial evolution steps. The time-like segments are implemented by evolving the state with the time evolution operator, which can be naturally done by QIS tools, while the spatial segments are realized by gauge-string operators, or equivalently by the controlled insertions and motion of external gauge charges. To realize the PDF, two fermionic field operators are then inserted at the endpoints, as shown at the bottom of Fig.~\ref{fig:pdf_quantum} $(a)$.

The application of this construction in the Schwinger model is shown in panel $(b)$, using a TN approach~\cite{Banuls:2025wiq}. Here the target bound state is the first excited state, i.e. the vector meson discussed above, which in the massless fermionic limit, corresponds to a fermion-antifermion (bosonic) bound state. The resulting distribution has then the expected qualitative features: the distribution is concentrated around $\xi=0.5$, consistent with approximately equal momentum sharing, while negative $\xi$ encodes the antifermion contribution through $f(-\xi)=-\bar f(\xi)$ for $\xi>0$, consistent with a vector state. The different data sets demonstrate that the PDF extraction has numerically converged, providing a rigorous extraction. One should recall that although these distributions are useful analogues of QCD PDFs, they do not enter any factorization formulas, as such concepts are not realized in $1+1$-d. Nonetheless, the methods used here provide a first step towards the full determination of these objects from first principles in gauge theories, and have since been more widely explored~\cite{Lamm:2019uyc,Li:2021kcs,Li:2022lyt,Grieninger:2024cdl,
Li:2024nod,Grieninger:2024axp,Kang:2025xpz,
Chen:2025zeh,Grieninger:2025mbm,Li:2026thq,Zou:2026cfk}.

\begin{figure}[h!]
\centering
\includegraphics[width=1\linewidth]{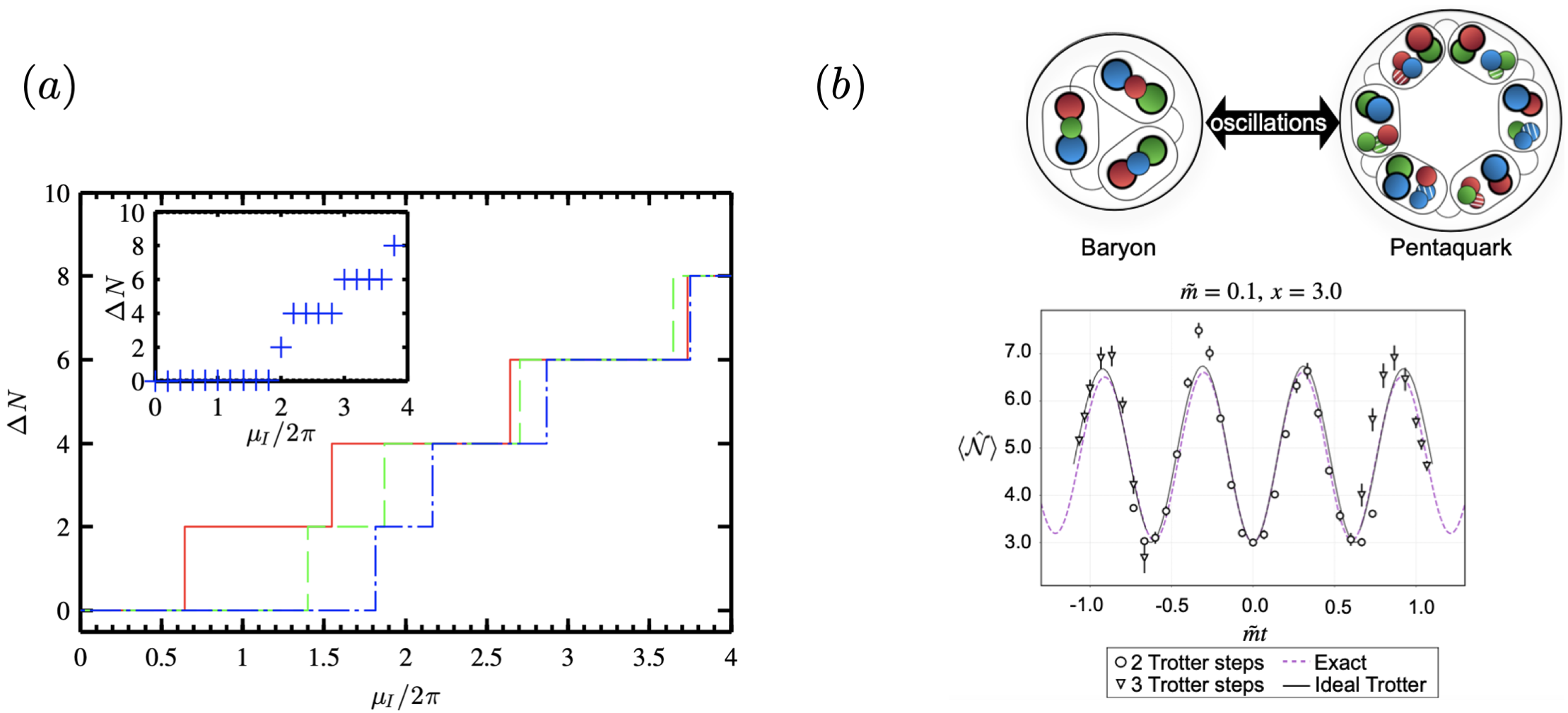}
\caption{Hamiltonian studies of finite-density gauge theories and
hadronic dynamics. $(a)$ Flavor imbalance $\Delta N$ as a function
of isospin chemical potential in the two-flavor Schwinger model,
using TNs. The discontinuities mark transitions between isospin sectors.
$(b)$ Real-time evolution in a one-dimensional SU(3) gauge theory,
initialized in a baryon state containing a static heavy diquark.
The total particle number tracks dynamics dominated by
baryon--pentaquark oscillations. Experimental results are compared
with exact and ideal (Trotter) evolution.
Figures adapted from~\cite{Banuls:2016gid,Atas:2022dqm}, respectively.}
\label{fig:finitemu}
\end{figure}

A closely related class of problems concerns gauge theories at non-zero
density, where conventional Euclidean Monte Carlo calculations can be
hindered by a sign problem. An example is shown in
Fig.~\ref{fig:finitemu} $(a)$, where matrix product states were used to
study the two-flavor Schwinger model at non-zero isospin chemical
potential~\cite{Banuls:2016gid}. The staircase structure of the flavor
imbalance $\Delta N$ reflects successive level crossings between
different isospin sectors, corresponding to first-order transitions.
The different curves show the dependence on spatial volume, while the
inset illustrates the transitions at finite lattice spacing. Quantum
devices also provide access to the real-time dynamics of hadronic
configurations. Panel $(b)$ shows an example in a one-dimensional
SU(3) gauge theory with light and static heavy
quarks~\cite{Atas:2022dqm}. Starting from a baryon containing two heavy
quarks and one light quark, pair creation generates additional
hadronic components, with the dynamics dominated by baryon--pentaquark
oscillations. These are tracked through the total particle number,
with the hardware results compared to exact and ideal Trotter
evolution. Together, these examples demonstrate access to
finite-density equilibrium properties and non-Abelian hadronic
dynamics, and we refer an interested reader to~\cite{Silvi:2016cas,Silvi:2019wnf,
Yamamoto:2021vxp,Florio:2023kel,Hayata:2023pkw,
Than:2024zaj,Fujikura:2026tqk} for related discussion.

\subsection{Hard probes of the QGP}\label{subsec:hard_probes}

In Section~\ref{subsec:HICs} we discussed how energetic or massive states formed in heavy-ion collisions could provide  a way to characterize the bulk matter produced in such events. Their production can be organized using the factorization picture introduced in Section~\ref{subsec:pp}, while their subsequent in-medium evolution is sensitive to real-time correlations of the medium. This separation makes them particularly natural targets for QIS methods: the hard probe can be viewed as a relatively small quantum system coupled to a complex many-body environment.

For a hard probe described by a density matrix $\rho_{\rm p}$, interacting with a medium state $\rho_{\rm med}$, the real-time evolution for an initially factorized probe--medium state can be written schematically as
\begin{equation}
\rho_{\rm p}(t)
=
{\rm Tr}_{\rm med}
\left[
U(t)
\left(
\rho_{\rm p}(0)\otimes \rho_{\rm med}(0)
\right)
U^\dagger(t)
\right],
\label{eq:hard-probe-reduced}
\end{equation}
where $U(t)$ is the unitary evolution generated by the full probe+medium Hamiltonian, and assuming an initial factorization of the full density matrix. Equation~\eqref{eq:hard-probe-reduced} is hard to evaluate as it requires evolving in real-time both the probe and the bulk matter. As a result, effective formulations of just the probe evolution in a simplified matter description are typically employed, see e.g.~\cite{Akamatsu:2014qsa,Brambilla:2016wgg,
Blaizot:2017ypk,Barata:2023uoi,Arleo:2026urv}. Even in such cases the resulting dynamics can be complex to evaluate numerically. As a result, QIS can be used both to provide novel methods to compute this reduced evolution or to extend it such that the dynamics of the bulk state can be captured.

\begin{figure}[h!]
\centering
\includegraphics[width=1\textwidth]{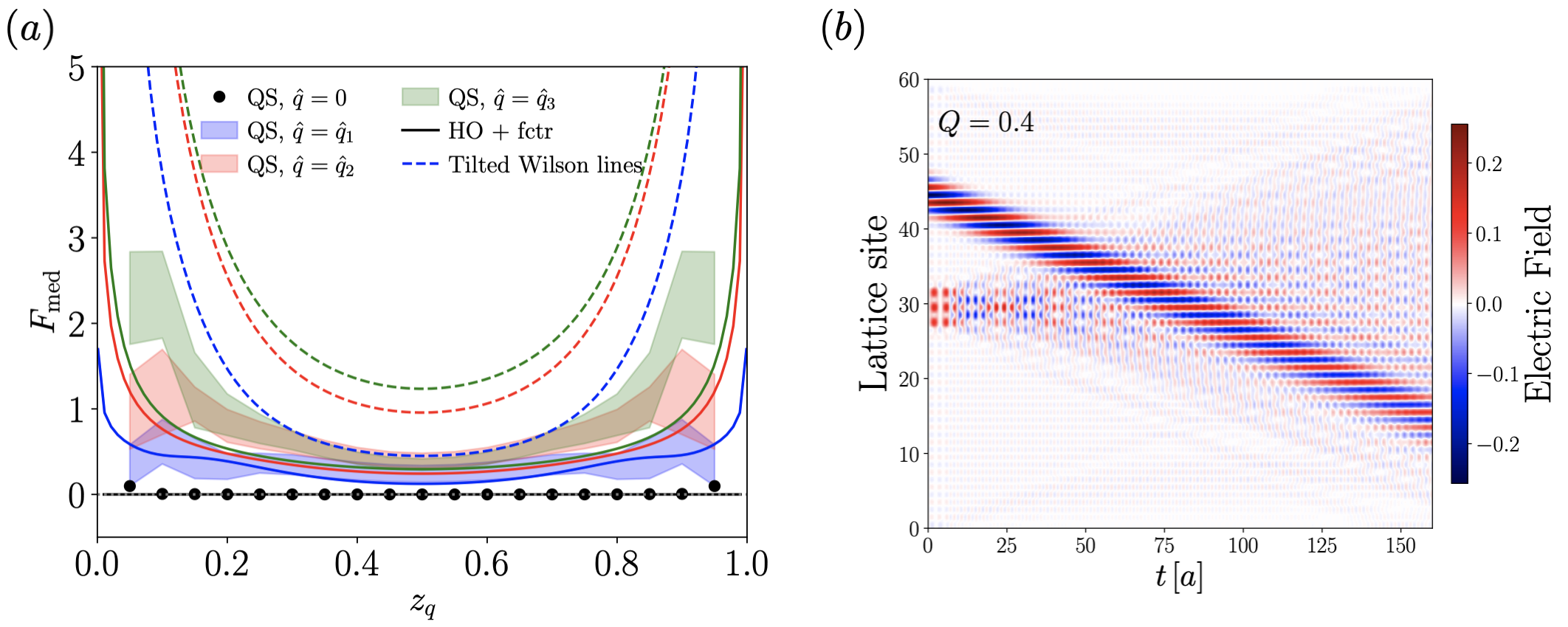}
\caption{QIS studies of jets in bulk matter.
$(a)$ Quantum simulation extraction of leading perturbative branching probability in matter. The observable $F_{\rm med}$ quantifies how efficiently the medium resolves the partonic substructure of the probe.
$(b)$ Real-time TN simulation of a probe scattering through a dense target in the massive Schwinger model, shown through the spacetime evolution of the electric field. Figures adapted from~\cite{Barata:2025hgx,Barata:2026icn}, respectively.}
\label{fig:jet_Qis}
\end{figure}

In Fig.~\ref{fig:jet_Qis} we show examples for both these points in the context of jet fragmentation in heavy-ion collisions. Panel $(a)$ shows the computation of the leading perturbative correction capturing the medium induced modifications to the partonic cascade coming from an initial energetic parton, as discussed above, and the quantity shown, $F_{\rm med}\propto dI/(d\omega d^2k)$, measures the degree to which the medium resolves a multi-parton system~\cite{Dominguez:2019ges}. The results compare a quantum circuit (QC) based formulation, evaluated through classical simulation, with approximate analytic results. The differences between the curves reflect the approximations in the analytic solutions, while the circuit formulation provides a numerical result within the specified theoretical approximation and discretization, comparable to state of the art phenomenology~\cite{Leitao:2026fgh}. The quantum circuit formulation  maps the relevant color and kinematic degrees of freedom to a quantum register, allowing the interference structure of the amplitude to be computed directly, see also~\cite{Barata:2021yri,Barata:2022wim,
Barata:2023clv,Qian:2024gph,Castro:2025ocx, Farrell:2024mgu,Li:2025sgo,Clemente:2022nll,Ramirez-Uribe:2021ubp} for related work.

Panel $(b)$ shows a similar realization of the same concept, but allowing for a dynamical background. A localized probe state is evolved through a dense target in the massive Schwinger model (propagating band), and the electric field is tracked as a function of space and time. At the same time, a \textit{dense} matter state is prepared at the center of the lattice, mimicking the bulk matter. The simulation captures essential features of jet evolution in heavy-ion collisions: the outgoing probe is modified by its interaction with a dynamical medium, and part of its energy is deposited into the target. This can be seen through a modification to the target and probe states over time, and is also reflected in other local observables or the entropy profile. This classical TN calculation extends the real-time study of probe--medium interactions in a controlled toy model to dynamical matter, and might in the future provide new clues towards better understanding emergent phenomena in such events.

\begin{figure}[h!]
\centering
\includegraphics[width=.9\textwidth]{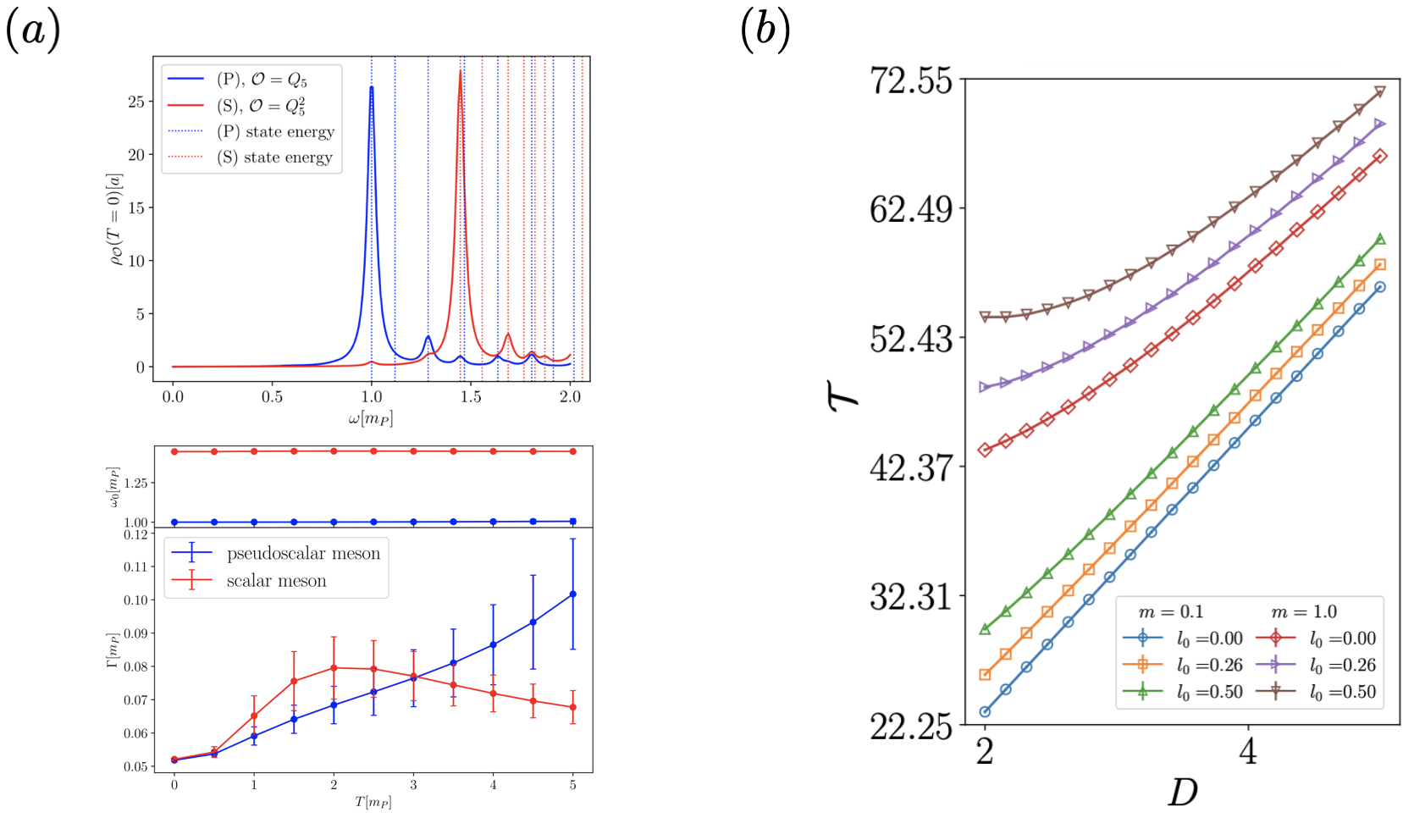}
\caption{QIS approaches to quarkonium probes.
$(a)$ Real-time spectral functions for mesonic excitations in a finite-temperature lattice gauge theory, illustrating how bound state peaks can shift, broaden, or disappear in a thermal medium.
$(b)$ Open system TN study of meson thermalization in the Schwinger model, where the thermalization time $\mathcal{T}$ is extracted as a function of the dissipator strength $D$.  Figures adapted from~\cite{Barata:2025jhd,Angelides:2025hjt}, respectively.}
\label{fig:quarkonia_Qis}
\end{figure}

As reviewed in Section~\ref{subsec:HICs}, heavy quark-antiquark pairs produced early in the collision and subsequently propagating through the QGP, also offer important probes of the bulk. The medium modifies their binding potential and can dissociate the bound state, leading to the sequential suppression pattern shown in Fig.~\ref{fig:jetsquarkonia_HIC}. In Fig.~\ref{fig:quarkonia_Qis} $(a)$ we illustrate how the spectral functions for the vector and scalar mesons in the massive Schwinger model are modified at finite temperature, extracted using exact diagonalization techniques~\cite{Barata:2025jhd}. Peaks in the spectral function identify bound state like excitations, while their thermal modification gives access to mass shifts and broadening. Real-time methods provide direct access to these quantities, while Euclidean approaches require spectral reconstruction through an ill-posed inverse problem. The vector and scalar operators couple to the corresponding states, and the fitted peaks broaden with temperature, qualitatively analogous to thermal broadening in QCD~\cite{Laine:2006ns,Brambilla:2008cx,Rothkopf:2011db}.  In panel $(b)$ an open quantum system formulation for meson evolution in a thermal environment is studied as a toy model for quarkonium dynamics. A mesonic excitation, represented by the vector boson in the Schwinger model is coupled to a thermal environment, and its thermalization time $\mathcal{T}$ is extracted as a function of the dissipator strength $D$ and other microscopic parameters. In this formulation one can separately track local observables, the decay of the electric flux string, and the mutual information between the two halves of the meson, providing a detailed diagnostic of the bound state dynamics. See also~\cite{Kim:2026cyu,Sharma:2023dhj,
Miura:2019ssi,Brambilla:2020qwo,Du:2023ewh,Verde:2026hcg} for related studies.

\subsection{Out-of-equilibrium and thermalization dynamics}\label{subsec:lthermalization}
The properties of QCD matter can be further explored by directly simulating such complex states. QIS platforms offer a unique opportunity to directly observe how collective phenomena and emergent properties can appear starting from first principles quantum simulations, or by emulating the dynamics in the natural Hamiltonian of controllable devices. Such an opportunity is quite exciting, as much is still to be understood about the emergence of collective behavior and many-body phenomena in QCD, and thus any new tools can provide important insights.

\begin{figure}[h!]
\centering
\includegraphics[width=.8\textwidth]{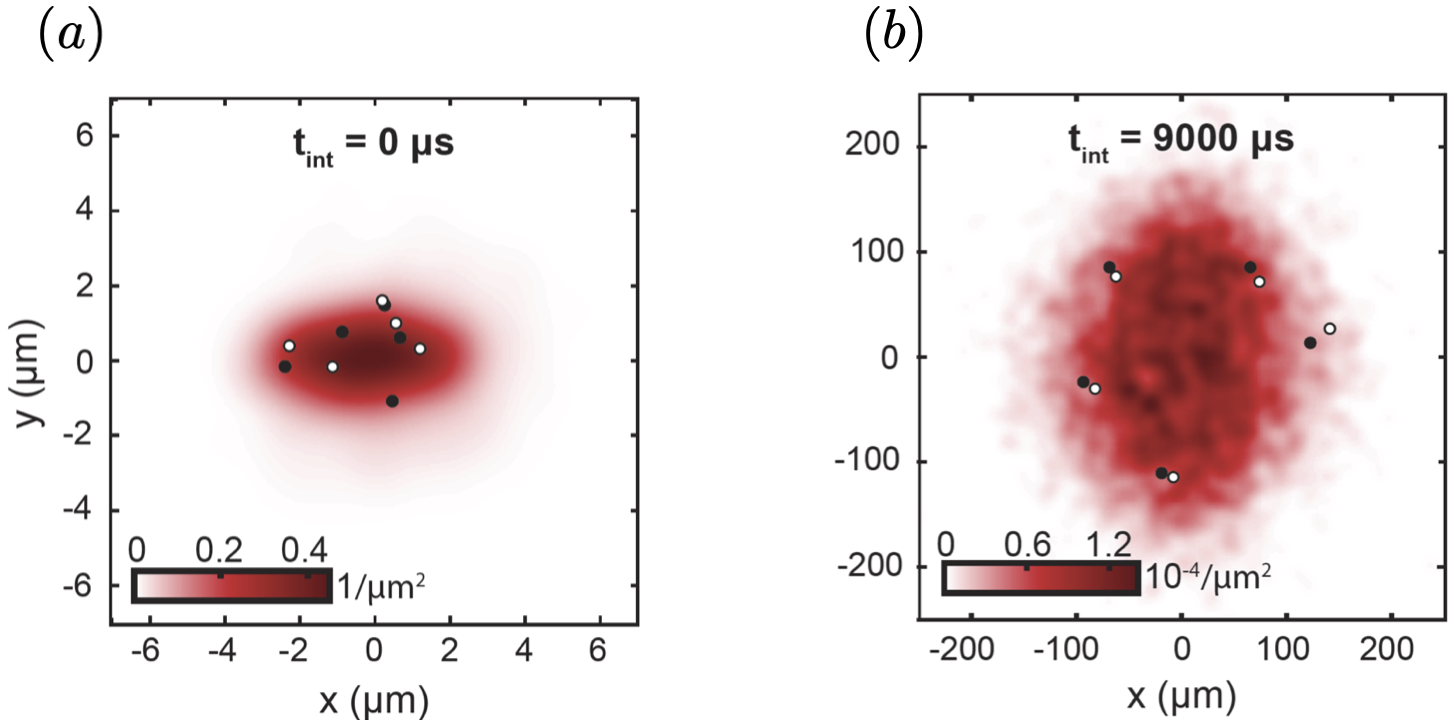}
\caption{Interaction driven expansion in a few body ultracold atom quantum simulator. An initially anisotropic cloud of strongly interacting fermionic atoms expands after release from the trap. The interactions convert the initial spatial deformation into a momentum space anisotropy, in close analogy with the mechanism that produces elliptic flow in heavy-ion collisions. Figure adapted from~\cite{Brandstetter:2023jsy}.}
\label{fig:fluffy}
\end{figure}

In this short section we provide two examples of this. In Fig.~\ref{fig:fluffy}, we show experimental results from an ultracold atom platform, where one can observe the dynamical response of a state made of fermions (dots) initially pinned in an anisotropic spatial configuration (panel $(a)$). At late times (panel $(b)$), one observes a spatial inversion of the particle distribution. This is an analog simulation of the collective expansion mechanism that also underlies the hydrodynamic interpretation of the bulk matter expansion in heavy-ion collisions, giving evidence for the existence of collective dynamics. We note that although this system is not 
a gauge theory, it allows to study how emergent collective behavior appears, with an exact control on the number of particles, see~\cite{OHara:2002pqs,Cao:2010wa,JosephThomasKulkarniAbanov2011} for related work, and~\cite{Janik:2025bbz,Caliz:2026znm} for recent studies on emergent hydrodynamical behavior from quantum simulation. The shape and dynamical response of the bulk can then be characterized in terms of the Fourier decomposition, as introduced above and done in heavy-ion phenomenology.

\begin{figure}[h!]
\centering
\includegraphics[width=1\textwidth]{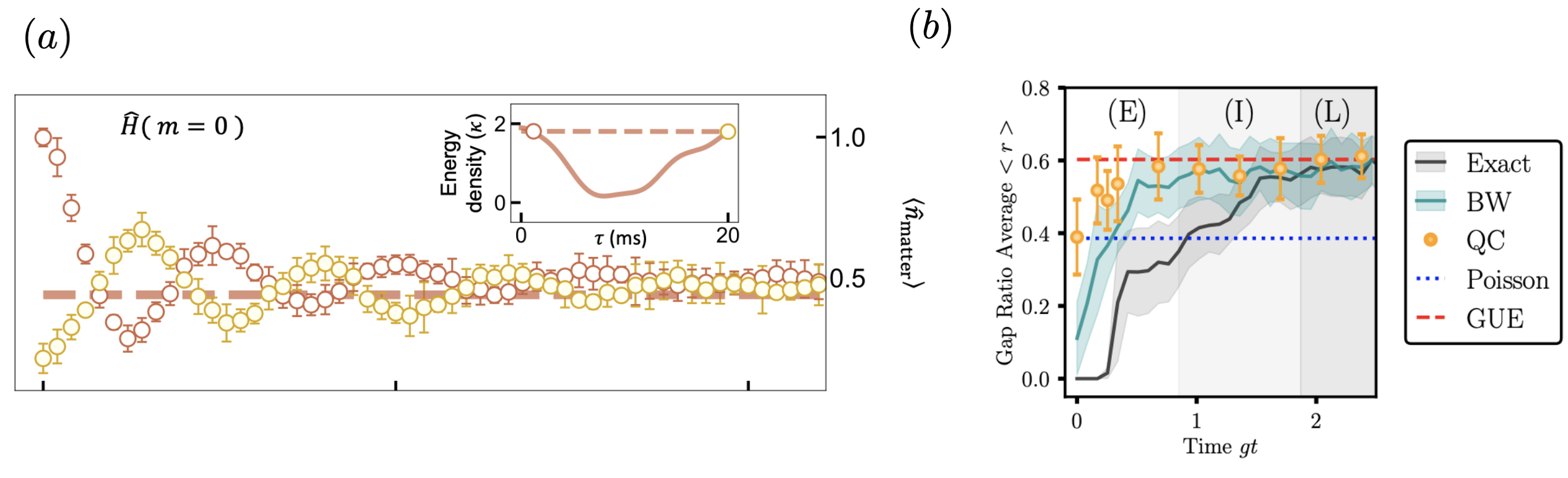}
\caption{Thermalization dynamics via  quantum simulation of gauge theories. The examples shown here probe how constrained gauge theory dynamics relaxes after a far-from-equilibrium quench. $(a)$ The analog U(1) simulator tracks the relaxation of local matter observables toward thermal behavior. The dashed line indicates the thermal equilibrium value, which is reached by different states with the same total energy. $(b)$ Trapped-ion $\mathbb{Z}_2$ simulation using entanglement diagnostics to identify early signatures of quantum chaos. Figures adapted from~\cite{Zhou:2021kdl,Mueller:2024mmk}, respectively.}
\label{fig:therma}
\end{figure}

Figure~\ref{fig:therma} exemplifies how quantum simulators can be further used to study the detailed dynamics of equilibration and thermalization of quantum systems. Although in heavy-ion collisions the initial state is far from equilibrium, the detailed mechanisms that drive the system towards equilibrium are only partially understood, with a complete first-principles description across the relevant regimes still lacking, see~\cite{Berges:2020fwq} for a recent review. In panel $(a)$ we show a U(1) lattice gauge theory simulated in a Bose Hubbard quantum simulator. The system is driven out-of-equilibrium by a global quench~\cite{Zhou:2021kdl}, and subsequently relaxes towards the value predicted by the appropriate thermal ensemble. Although the full state evolves unitarily, local observables can approach thermal values through dephasing and entanglement with the rest of the system, with quantum typicality and eigenstate thermalization hypothesis providing theoretical explanation for the thermalization process in a closed system, see~\cite{Deutsch:1991msp,Srednicki:1994mfb, Goldstein:2005aib,Popescu:2006rhr}. In panel $(b)$, a digital trapped-ion quantum computer is used to study a $\mathbb{Z}_2$ lattice gauge theory in $2+1$-d~\cite{Mueller:2024mmk}. Randomized measurements give access to the entanglement Hamiltonian, whose level statistics allow to diagnose the onset of quantum chaotic behavior. Similar studies of thermalization in a variety of related models have been also widely explored in recent years, see e.g.~\cite{Banuls:2010zki,Buyens:2016hhu,Hayata:2020xxm,
Yang:2019rtr,Delacretaz:2022ojg,Luo:2023rso,
Hayata:2024smx,Hayata:2026rmv,Cuntin:2026nlz}, while searches for thermalization breaking states in gauge theories have also attracted attention, see e.g.~\cite{Banuls:2010zki,James:2018qly,Robinson:2018wbx,
Banerjee:2020tgz,Halimeh:2022rwu,Desaules:2022ibp,
Desaules:2022kse,Ge:2023xkf,Sau:2023clm,
Hayata:2023puo,Budde:2024rql,Calajo:2024bvs,Barata:2026scx}.

\section{Conclusion}\label{sec:conclusion}
In this article we have discussed how quantum information science
can help address open questions in high energy nuclear physics.
Its relevance is particularly clear when the physics involves
real-time evolution and strong quantum correlations, which remain
challenging for much of the established QCD toolkit. Tensor networks
and quantum simulation provide new ways to investigate these regimes,
while quantum information concepts offer additional insight into
the quantum dynamics.

The applications reviewed here mostly connect these methods to the physics
of high energy collisions, from the formation and internal structure
of hadrons to their interactions with QCD matter and the emergence
of collective behavior. Although most studies still rely on simplified
theories or small computational systems, they already provide controlled access
to mechanisms that are difficult to study otherwise. Building on
these results toward QCD will require larger simulations and more
realistic theoretical constructions, together with systematic control of numerical
and experimental uncertainties.

Several relevant directions lie beyond the scope of this review, such as
the collective neutrino flavor evolution~\cite{Hall:2021rbv,
Arguelles:2019phs,Roggero:2021asb,Roggero:2022hpy,
Amitrano:2022yyn,Chernyshev:2024pqy},
the vacuum and low-energy structure of field theories~\cite{
Dempsey:2023gib,Rogerson:2026jjt,Dempsey:2022nys,
Chakraborty:2020uhf},
and chiral and anomalous dynamics~\cite{Kharzeev:2020kgc,
Ikeda:2023vfk,Ikeda:2024rzv,Okuda:2024jzh}, among several other topics.
More experimentally driven areas of research in the intersection of NP and QIS have also gained traction in the past few years, the most explored one concerning the measurement and extraction of quantum correlations at colliders,
from top-quark entanglement~\cite{Afik:2020onf,Dong:2023xiw,
ATLAS:2023fsd,CMS:2024pts}
to hyperon spin correlations and their connection to
hadronization~\cite{Gong:2021bcp,Barata:2023jgd,
BESIII:2025vsr,STAR:2025njp,CMS:2026edg,
Barata:2026hck,Liu:2026ees}.
These topics illustrate that QIS also offers new formulations of
physical observables and new ways to understand quantum
correlations in high energy systems.

Progress toward applications with direct phenomenological relevance
will require advances in both hardware and algorithms, together with
careful choices of observables and target problems of interest. The results reviewed in this article lay the foundations for a new paradigm in nuclear physics, in which quantum information provides both a computational framework for studying dynamics otherwise  difficult to access with established methods, and a language for formulating new physical questions in nuclear physics.

\section*{Declaration of generative AI and AI-assisted technologies in the writing process}

During the preparation of this chapter, the author used ChatGPT (OpenAI, Pro subscription) to assist with editing and reviewing the author’s original draft for language, clarity, and readability. The author reviewed and revised all suggested changes and takes full responsibility for the content of the published work.

\bibliographystyle{elsarticle-num}
\bibliography{reference}

\end{document}